# ANALYTICAL MODELS OF FOREST DYNAMICS

## *Georgy P. Karev*

NCBI, NIH, 8600 Rockville Pike, Bethesda, MD 20894, USA

## ABSTRACT

The hierarchical system of forest ecosystem models based on the theory of individual-based (structured) models of populations and communities is briefly described. New self-thinning models are integrated with tree stand models within a hierarchical system of models, aimed at assessing (quasi-) stable states of a forest ecosystem under given environment and management.

Within the structural model of succession, the "ergodic theorems in biology" (announced earlier) are proved. They assert that the areas of biocenoses in a climax state of association should be proportional to the "proper times" of their development in a succession line; the other characteristics of biocenoses should be also proportional to their own "proper regeneration times" in a climax state.

A simple theoretical approach to estimating the deviation of the current state of forest ecosystem from its steady state is derived and applied to data from the Prioksko-Terrasny Reserve. The results obtained can serve as a theoretical basis in numerical processing of data on ecological monitoring of undisturbed forest lands.

## 1. INTRODUCTION

*The individual-based approach* is currently an intensively developing approach in the modelling of various ecological systems (Levin, Hallam and Gross, eds., 1989; DeAngelis, Gross, eds, 1991; Antonovsky et al., 1991; Huston, 1991; Judson, 1994; Liu, Ashton, 1995; etc.). Individual-based or structural models are adequate tools for linking behaviour of the population as a whole with dynamics of individuals and their interactions. The models represent the dynamics of the ensemble of trajectories of the individuals. The analytical models analyse the evolution and asymptotic behaviour of the initial distribution of individuals; the simulation models calculate directly the dynamics of each individual or some integrated unit, such as an age cohort. Gap-modelling of forest ecosystems is one of the main fields of application of both simulation (Botkin 1993; Shugart 1992, 2004) and analytical (Karev 1995; Chertov, Komarov and Karev 1999, ch.3) individual-based models.

The analytical models compose natural part of biologically-based process (or "mechanistic") models intended for understanding of forest dynamics. There exist some known contrasts between biologically-based process models and management-oriented models (Mohren, Burkhart, 1994). Nevertheless, we will show (s.4) that the analytical model, in particular structural models of successions could be applied to estimating of long-term forest conditions under given management and so may be a useful link between both mentioned classes of forest models.

Structured models of metapopulation form one of the new branches of mathematical modelling of biological populations and communities. This theory was developing in two essentially different directions: modelling of moving individual metapopulations (Gyllenberg, Hanski 1992; Gyllenberg, Hanski, Hustings 1997) and modelling of metapopulations of motionless (fixed) individual, namely, plant metapopulations and forest communities (Karev 1994; Liu, Ashton 1995, etc). (It seems to be possible and desirable the synthesis of these two directions; the natural object for the synthetic mixed model may be the «forest - insect» ecosystem, see, e.g., Sukhovolskii, 1996).

The results of the metapopulation approach can be considered as a newly formulated theory of structural modelling of forest community dynamics, based on the following components: a) the layer-mosaic concept of spatial-age structures of forest communities, the so-called "gap paradigm"; b) computer simulated gap-models; and c) the theory of structural models of populations and communities. The main goal of this theory is the developing of methods of construction and investigation of hierarchical systems of models based on different approaches and in various spatial and temporal scales. The aim of such hierarchical system is not to include each and every detail, but to link models across scales, to calibrate more aggregate models using detailed analysis at a smaller scale, and also to obtain consistency in the analysis of ecosystem dynamics in a hierarchical approach.

The hierarchical system of models may be useful both for computer simulations and analytical modelling although the aims of these two approaches to studying of forest ecosystem dynamics are different. The main aids and advantage of analytical modelling is the investigation of the asymptotical behaviour of the system dynamics, both in space and time. The time asymptotic of the system conditions shows the direction of the system development and its final condition. Such investigation is mostly useful under constant environment and management or, more generally, with management, which may depend on the system condition but do not depend explicitly on time. It allows estimating the final result of any particular management under constant environment of the system.

Practically we could use these estimates for preliminary studying, which allow choosing the acceptable or the most desirable variants of management and final conditions of the ecosystem according particular (e.g., ecological or productive) criteria. The detailed investigation of the chosen variants should be done with the help of computer simulations using the same (or close) hierarchical system of models. From theoretical point of view, the system of analytical models together with mathematical methods of their asymptotical investigations composes an important part (together with large and detailed models for computer simulations) of the general theory of forest ecosystem dynamics, which hopefully will be created in the nearest future.

## 2. THE HIERARCHICAL SYSTEM OF MATHEMATICAL MODELS OF UNTOUCHED FOREST COMMUNITIES

The forest community is very complex. According to the general principles of system analysis, it is necessary to construct not one model, but a hierarchical set of models for the mathematical description of the structure and dynamics of these complex systems, such that a model at a particular scale includes models of the previous scale as elementary objects or ready-made modules. The principles of construction of such a hierarchical system can be different, for example, with respect to spatial scale, significant time-periods of the dynamics, or levels of structural organisation.

It is possible to consider a number of mathematical models of forest communities as components of the hierarchical system of models of tree populations and communities, constructed on the basis of increasing complexity of structural organisation and corresponding expansion of spatial and temporal scales, namely: «a single tree - locus (or gap) - population - meta-population- community or forest landscape». Consider briefly the basic stages of this system; a more detail description is given in (Chertov et al. 1999; Karev 1999).

## 2.1. Free-Growing Tree Models

The initial item in this system is a model of a free-growing tree. There are a large number of different models of a single tree or plant (see surveys Thorhley 1976; Antonovsky et al. 1991; Kull, Kull 1989; Korzukhin, Semevsky 1992; etc). Very different particular models (Poletaev, 1979; Karev, 1984, 1985; Galitsky, Komarov, 1987; Oja, 1985, 1986, etc.) or any model of tree growth used in gap-models (e.g., JABOWA, FORSKA etc, see Botkin, 1993; Shugart, 2004) may be taken as appropriate examples of free- growing models of various complexities.

## 2.2. Modelling of Tree Competition

Mathematical modelling of interactions and competitions between separate trees for external resources (light, water, area of growth, etc.) is a necessary link to proceed from the models describing separate trees to the models of the population level.

The theory of plant competitions is an important part of plant ecology (Uranov, 1965; Thorhley, 1976; Wu et al, 1985; Walker et al, 1986; Ford, Sorrensen, 1991; Mou et al., 1993; Li et al., 2000,etc.). Application of competition indices associated with the «nutrition area» (polygon models, growth zones, etc.) is of little use in this case. The most complete mathematical description of the competition processes, by taking into account the spatial structure, is given by the interaction potentials with the subsequent including into the model a forest stand describing with the aid of Markov' fields or distributed integro-differential equations (Komarov, 1979; Tuzinkevich 1988; Grabarnik 1992). However, such an approach entails significant mathematical difficulties, which are not always adequate to the tasks to be solved.

A more widespread approach is to describe interaction processes by calculating the external resources available to an individual tree. The results obtained are then described with the aid of the «regulating functionals» or briefly «regulator» (Tucker, Zimmerman, 1988) that represent certain average values for the current distribution of the trees in the population. Typical examples of this approach are models describing competition for light (the leaf canopy usually being simulated using the random turbid medium concept (e.g., Monsi, Saeki, 1955; Kull, Kull, 1988; Korzukhin, Semevsky,1992) with some modifications for poly-dominant and multi-layer stands (e.g., Karev, 1984, 1985); models analysing competition for area of growth (e.g., Galitsky, Komarov, 1987); distributed models of root competition (e.g., Karev, Treskov, 1982) and other more or less complex models (Pacala, Silander, 1985; Post and Pastor, 1990, 1996, etc.). A theory of plant competitions applicable either for analytical or computer models was developed in (Adler 1996).

## 2.3. Models of Tree Subpopulations

The transition up to subpopulation scale (loci, cenon, gap - see definition below in n.4.2.5) is made by synthesis of the free- growing tree model together with models of competition for external resources, such as sub-models of soil organic matter dynamics and competition for light. A model of the dynamics of single generation of a tree sub-population was the result of the synthesis e.g., the various separate gap models.

The "dynamic models of tree growth" DMTG (see s. 3.3 below) is an example of more detail subpopulation model, which is intended for description of dynamics of the main characteristics of tree stands: number of trees; average height; average diameter; average volume of single tree; growing stock; wood litter. Data on trial areas of Scots pine and Norway spruce stands (Advance…1964), data on Scots pine plantations of various densities (Rubzov et al., 1976), and also growth yield tables of "normal forest" (Zagreev et al. 1992, Kozlovskij, Pavlov, 1967) of various species, were used for verification of the DMTG model, which show a good accuracy.

## 2.4. Some Classes of Tree Population Models

Taking account of interactions amongst trees, described by appropriate regulators, complicates the investigation and application of models. These difficulties may be overcome in some classes of models of theoretical and practical interest, which are defined by additional conditions of the initial or current distribution of sub-populations: 1) a single-species, even-aged tree stand (a sub-population with uniform distribution of size and species); 2) a multi-layered tree stand (i.e., non-uniform distribution, e.g., a population divided into m groups with identical trees within each group); 3) a mixed stand (practically the same as a model of a many-layered tree stand but the equations describing dynamics of the *i*-th species can depend on its number); 4) a tree stand with a *narrow* current distribution of sub populations (with moments greater than 2 equal to zero); and 5) a tree stand with a narrow initial distribution. Models 1) to 3) can be used when the initial distribution can be approximated by the piecewise constant function, whereas models 4) and 5) are especially useful for modelling gap dynamics construction. The essential feature of models 4) and 5) lies in dividing the initial model into two parts. The first part allows regulators to be taken into account and gives a result that is used to describe the dynamics of a separate gap in the second part of the model. Remark that the first stage model really describes characteristics of the main population dynamics and is interesting itself. The methods of construction of these models, with specific examples, were given in (Karev 1995, 1999).

## 2.5. The Layer-Mosaic Concept and «Elementary Units» of Forest

The layer-mosaic concept (or gap paradigm) considers a forest community as a meta-population consisting of a large number of tree subpopulations (loci or gaps). The initial idea of the concept, from the beginning of the century, is that the «forest elementary unit» is not an individual tree but an association of trees. The forest community is considered to be a system of spatial mosaics consisting of asynchronously developing patches, at different stages of development and changing in time as a result of internal dynamic processes.

It seems that the layer-mosaic concept, or "gap-paradigm" arose after work (Watt 1947; Dylis 1969; Aubreville 1971) and it has developed intensively in last years. A detailed account of the concept was made, for example, in a monograph (Popadyuk et al. 1994); authors analysed its application to deciduous forests consisting of mono- or poly-dominant associations of trees, named "loci". A locus is determined by a definite species structure (i.e., density, age, sizes, etc.) at a particular time. The features of development and processes of the formation of loci as patches in «geographical» space were investigated in the monograph. A verbal model was constructed and then implemented as a detailed computer simulation model.

Another verbal model put forward in (Buzykin et al. 1985, 1987), was concerned with research about characteristics of the limiting mode (i.e., whether the regime is stationary, oscillatory or chaotic), which is established in a forest community in the absence of external destabilising factors. This question is one of the basic problems of the study of biological communities and their models. In these papers, the concept of the «cenon» as an elementary structural unit of the forest community was defined (by obvious physical analogy). The cenon is a part of the community of forest plants characterised by uniformity of the area occupied and by uniformity of the generation stage. The patch (or cenon) structure of tree populations was demonstrated and its endogenous dynamics were described, with Siberian spruce as an example.

The basic ideas may be summarised as the following "axioms of the cenon concept ":

1. The forest community can be considered as a meta-population of cenons.
2. Each cenon is a sub-population of trees of one generation, which occupies a fixed area, and arises, develops and perishes as one unit.
3. The birth of a new cenon in the area occurs only after destruction of the old cenon occupying it.
4. The growth of an individual tree depends on interactions with other trees in the cenon, and does not depend on trees belonging to other cenons.

From these, some fundamentally important conclusions for modelling arise:

(a). the model of the forest community considers a cenon (locus or gap) as a new individual object, rather than a separate tree;
(b). the model of an individual cenon coincides with a model of one generation of a tree population;

(c). the dynamics of a meta-population of non-interacting cenons is described by an autonomic structural model; asymptotic behaviour of such models was completely investigated (Karev 1993).

The most known kind of forest «elementary unit» is «gap» (Botkin 1993). A model of separate gap describing the dynamics of trees on sites of specified area is the basis of every gap-model. Each tree of a given species is characterised at any one time by a certain set of variables: height, diameter, etc. The equation for growth depends on light, temperature and other environmental parameters; competition for resources can also be included. In the vast majority of gap models, the influence of other gaps is not taken into account. Tree supplement and mortality on a site is described by stochastic processes; the initial distribution of trees is also stochastic. As the model is stochastic, forecasts are calculated as the average of a rather large number (80-100) of independent solutions.

Although the concepts of forest elementary units (gap, locus and cenon) are rather similar, there are qualitative differences between them. Gap is a conditional "topographical" cell, i.e., the site of the fixed area in which the birth, growth and death of each separate tree is simulated; locus is a real forest structural unit, and its spatial allocation in "topographical" space reflects actual complex processes of renewal and development of forest cover in the released site; cenon is a cell not only in "topographical" space but also in "phase" space of forest cover. This means that the trees composing a cenon form a rather homogeneous sub-population, so that the degree of uniformity increases with age and the interactions between these trees are essentially stronger than those between trees belonging to different cenons. The reasons for cenon formation in "topographical" space are the same as for locus formation. The allocation of a cenon as an uniform formation in phase space is caused by the action of «discriminate decay», consisting of sharply increasing mortality of trees with a diameter less then the average in a sub-population. This phenomenon, resulting in formation of a homogeneous sub-population (a cenon), was demonstrated in the works (Buzykin et al. 1985, 1987). The appropriate mathematical models of cenon formation were constructed in (Karev 2001).

## 2.6. Gap-Models and the Problem of Expansion of Spatial and Temporal Scales

Gap-models have been widely used in the years since the first JABOWA model. Monographs (Botkin 1993; Shugart 1992, 2004) contain the basic concepts, models and results of gap modelling.

From the standpoint of theoretical ecology, gap-models represent realisations (more or less simplified) of the layer-mosaic concept. Models of the latest generation (FORET, FORSKA, et al.) are distinguished by a considerably more extensive use of theoretical ecology to justify them, as compared with the first gap-models. On the other hand, gap modelling may be considered as a large-scale computer experiment to verify the main principles and consequences of the gap-paradigm.

Some gap-models (e.g., ZELIG) take into account processes of interaction with trees in the neighbouring patches (e.g., competition for light). The majority of models assume that there are no significant interactions between trees belonging to different gaps. Calculations with the appropriate models confirm this assumption, namely: 1) attainment of satisfactory forecasts of the dynamics of various forest communities, from boreal to tropical, when no account is taken of interactions between gaps; 2) the absence of interactions leading to rapid movement of the gap ensemble distribution towards a stable situation; this convergence can be observed in computer experiments with the model and is supported by analytical models (see below). On the other hand, a sufficiently strong interaction between patches may force synchronisation of their development. Thus the system as a whole may have an oscillatory mode not observed in natural forests, which are not subject to broad-scale influences.

So the assumption about absence of essential interaction between patches can be considered as acceptable first approach. This assumption is very important for both computer and analytical study of gap-models in large spatial scales.

Research on the influence and long-term consequences of global climate change and anthropogenic impacts on forest vegetation requires a transition to long time scales and should include study of asymptotic behaviour of the model with various environmental scenarios. The problem of the necessary enlargement of the spatial and temporal scales of local gap-models is a very important and difficult one. Reviews of relevant results have been published (Urban, Smith, 1989; Urban et al., 1991; Shugart, 1992; Acevedo et al., 1995; Karev, 1999; Chertov et al., 1999, etc.).

The enlargement of spatial and temporal scales conflicts with the individual orientation of gap-models that trace the dynamics of each separate gap (and even of each separate tree in a gap), so that the spatial and temporal scales of the model are limited by the capability of even the modern computer. Enlarging the time scale to take the dynamics of a large number of generations into account gives rise to the following problem. The output of the landscape gap-model is a map of a forest ecosystem represented by a mosaic of gaps, with certain co-ordinates and a range of states. The stochastic character of the processes of tree birth and death, with a random initial distribution, leads to the result that the observed and calculated states for each separate gap will be quite different (and so will the map as a whole), as a consequence of the stochasticity of the process. Thus, the major advantage of gap-models - "individual orientation" - is also a major constraint to enlargement of the spatial and temporal scales. On the other hand, the statistical characteristics of the map as a whole, e.g., age distribution, average height and diameter of trees, standing crop of wood etc., can be close to the real values.

These problems cannot be solved completely by using better programmes and more powerful computers, but only through appropriate mathematical approaches and methods. Some such methods have been based on the theory of structural models of forest dynamics, an approach that may be considered as a mathematical theory of gap modeling.

The research of asymptotic behaviour of large simulation gap-models seems to be impossible without inclusion them in frameworks of the analytical theory. Such inclusion can be carried out on the basis of the following thesis:

*gap - models can be considered as computer realisations of the structural models of metapopulations and communities.*

## 2.7. Meta-Population Approach for Enlarging the Spatial and Time Scales of Forest Models

Structural models of tree populations, in which the individual is a separate tree, have been put forward repeatedly, but the complex competitive interactions between trees have resulted in essential non-linearity of such models. Thus these models do not permit qualitative solutions and are even difficult for computer implementation.

The situation essentially varies with transition to the new individual object, the sub-population of trees (i.e., locus, cenon or gap), and consideration of a tree population as a meta-population of cenons or gaps according to the layer-mosaic concept, or gap-paradigm. The gap-paradigm allows the next (most important) step in enlarging the spatial and time scales of the model to be taken.

The first stage of spatial scale enlargement of gap-models is the transition from a model of a homogeneous tree sub-population (i.e., in a gap) to a model of a heterogeneous sub-population. The basic difficulty is to take into account of the processes of interactions amongst trees, i.e., dependence of the growth equations on regulating functions. Sub-population models of one generation may overcome this mathematical difficulty with an appropriate choice of initial distribution, or some of the assumptions about distribution of tree properties mentioned above in 2.4 (see Karev, 1995, 1999 for details).

In the models constructed by these methods, the interactions amongst trees turned out to be confined within a gap, so that the gap models can be rather complex. However, the structural model of a gap meta-population proved to be an autonomous one, on the basis of the assumption that it is possible to neglect any interactions between gaps. This circumstance is important, as it enables the asymptotic behaviour of all structural autonomous models and some of their generalisations to be investigated completely (Karev, 1993).

The second stage of spatial scale enlargement of forest community models is concerned with the modelling of meta-populations consisting of large numbers of gaps. Dynamics of a gap meta-population can also be considered on different temporal scales (i.e., over one or many generations). We emphasize that the model of one generation of a gap meta-population has the same formal structure as a model of one generation of a tree population. Models of one generation of gap meta-populations can be constructed by the same methods as sub-population models of classes 1) to 5) in 2.4.

For realisation of this approach and constructing the gap metapopulation model, it is necessary to find rates of birth and disappearance of individual gaps as an integrated process. The completion of this task essentially depends on the particular gap model. It appears that such fundamental characteristics as probability of disappearance, duration of gap life, and also life duration of a whole meta-population, depend on initial gap size and other characteristics of tree distribution (e.g., variance). These results again emphasise the critical importance of gap size, as has been pointed out both in theoretical work and in experiments with computer gap-models. Furthermore, the various classes of meta-population models may be constructed by the same methods as were used for construction of models of non-uniform sub-populations, if "tree" is replaced by "gap". Models of non-uniform meta-populations (including spatially non-uniform ones) are the most interesting (Karev 1995).

These models allow the study of dynamics of spatially non-uniform tree populations, remaining within the framework of the ordinary differential equation based models (although much more complex models based on partial differential equations are usually applied to such situations). The models of this class represent an adequate mathematical tool enabling the construction of models of meta-populations (i.e., forest communities) on large spatial scales on the basis of "local" models (i.e., gap-models).

At the third stage of spatial-temporal scale enlargement, it is possible to study asymptotic model behaviour over long time intervals. As all interactions between trees take place within a gap, the appropriate structural model of meta-populations turns out to be autonomous.

The general result is that, irrespective of the details of the description of the dynamics of an individual tree, and of the processes of competition and interaction of trees within sub-populations, there is a unique stable distribution for the meta-population as a whole, which is attained at an exponential rate, while originating from any initial state. The precise form of this limit distribution has been found dependently on the initial structural values, birth and death rates of sub-populations, as well as on the parameters, which could describe the environmental conditions (Karev 1993, 1999).

The limiting distribution describes an unique stable state of dynamic balance, representing a mosaic of gaps of various ages and states, which are formed, develop and disappear according to internal laws, but the distribution of all the sets of gaps turns out to be stationary. Obviously, this statement does not contradict results, which indicate that an oscillatory mode in a tree population may be established as a consequence of an oscillatory regime within a sub-population that is an element of a wider meta-population.

In many cases, exact expressions of variables for the limit distribution are not required, since only the basic, general variables, such as growing stock or height, are important. Average values of these variables at the limit distribution are given by the regulating functions that, in a limiting stable state, have been defined in precise form.

Earlier results have led to new methods of applying gap-models over long intervals of time and making analytical studies, using these models, of the influence of different environmental variables Using the analytical kernel of a computer gap-model (i.e., the equations, which describe the tree growth and number dynamics together with rules for birth and death of gaps and their initial distribution), it is possible to define the limit distribution of a meta-population and to find limit values of the necessary regulators. Moreover, it is possible to investigate the dependence of the asymptotic distributions on various parameters, included in the model (e.g., the initial distributions, etc.) by utilising various scenarios of climatic and anthropogenic influences. This research can be done, using available software for qualitative analysis of systems of differential equations.

## 2.8. Dynamics of Forest Areas and Structural Model of Succession

The top level of structural organisation and spatial-temporal scales includes the models that describe succession processes. The above approach describes asymptotic behaviour of the forest meta-population, making two assumptions: 1) that the forest vegetation already has a mosaic structure in its initial state; and 2) all gaps belong to a single type of forest cover, which is defined by species structure and site conditions.

Both assumptions appear too restrictive. Thus, if the initial structure of vegetation was fairly homogeneous (for example, after rapid colonisation of an area, as a result of catastrophic external influences, such as fire, insect attack, felling, etc.), the formation of small-scale structure occurs only after several generations. Therefore the dynamics of the forest vegetation can be simulated on larger spatial scales for areas, which are subject to strong external influences for about the duration of one generation. This aim can also be considered within the framework of structural models of successions and the dynamics of forest areas.

The term succession implies a study of changes in natural systems together with the causes and directions of such changes. There is considerable confusion between a precise meaning of this term and mechanisms constituting the basis for the process succession. Not trying to discuss at length the succession theory and modern attempts to reformulate it, some basic notions will be touched on below.

The natural dynamics of biotic communities give rise to consecutive change of ecosystems (biocenoses, phytocenoses), successively arising on a certain territory. This process is referred to as succession and a set of biocenoses involved in it is called a succession system.

An example of such a succession system can be found in a vegetative association, i.e. a set of comparatively uniform phytocenoses being on different stages of succession development. In the long run this set tends to a climax, i.e. a relatively stable state for a set of biocenoses replacing each other as a result of the succession process.

The concept of climax of a vegetative association allows one to predict the direction of a natural course of vegetation changes. The statement about uniqueness of climax is considered in some works as a natural consequence of deterministic development in a given environment. Thus, the different variants of climax can arise only by virtue of the environment change.

It is known, however, that the real sequence of changing and replacing biocenoses may be multivariant; some examples were given elsewhere (Reymers 1994; Razumovsky 1981; etc.). Moreover, the lack of uniqueness, the principally ambiguous dependence between vegetation in equilibrium and environment was revealed (Vedushkin et al. 1995)

Formally, if the transition from one succession stage to another is univalent and there are no close paths (cycles) under sequential transitions, then there exists evidently a single climax state. But the uniqueness of the climax is not obvious under multivariate transitions; moreover if any cycles attend the sequential transitions, then the concept of the climax state might be not entirely correct.

Thus, it is appropriate from the point of view of mathematical modelling to go over from the concept of climax state of a succession system to that of a stable or equilibrium state of corresponding mathematical model. Notice that this approach was actually used already in the first works on Markovian succession models. The practical importance for modelling of the choice of one or another concept of climax state will be shown in s.4.2.

Investigations of succession systems were mainly devoted to studying the types of succession and conditions of their phase changing. Qualitative and especially quantitative characteristics of climax (equilibrium) states were studied comparatively worse. Nevertheless it is necessary to describe quantitative characteristics of climax states for developing methods, which permit us to evaluate the deviation of observed ecosystem states from those in equilibrium. These methods of monitoring a steady-state succession system are especially important under changes of environment or for reserve lands.

Mathematical modelling of natural forest dynamics and succession was developing for the last 25 years in two essentially different directions (Acevedo, Urban and Shugart 1996; Liu and Ashton 1995; Botkin 1993; Shugart 2004, 1992). The first one is known as a succession gap-modelling, which was culminated in the models of JABOVA- type simulators. Alternative models are known as Markovian or transition models. Mathematical foundation of these two types of models is quite different.

Markovian succession models operate with proportions of territories occupied by successions residing at certain stages of development, without taking into account the intra-stage dynamics (Horn 1975; Shugart, West 1981; Cherkashin 1981). From mathematical point of view, the models of this type reduce to discrete Markov chains with a finite number of states which comply with different succession stages and whose transition probabilities vary in inverse proportion to the stage duration. Markovian models could be applied for relatively species-poor forests (i.e., boreal forests) or for cases when the detail dynamics is not of interest. The asymptotic behaviour of such models is well known. The only stationary state is established where stage areas are proportional to the components of eigenvector of stochastic transition matrix, beginning from any initial state. Some modern applications of Markovian models for natural course of succession through forest types was developed in a case study of mixed boreal forest in Prioksko-Terrasnyi Biosphere Reserve in (Logofet, Korotkov 2002).

Modelling of succession processes requires constructing gap models of a landscape level (Urban, Smith 1989; Urban et al. 1991; Acevedo et al. 1995, 1996) and "running" them on rather large intervals of time (corresponding to several hundred years) under different environment scenarios. So a spatial-time scale of the gap-model of succession covering a range of spatial and temporal scales could be limited, moreover, too many details may not allow one "to see forest behind trees". More essential problem is that any computer simulation gives only one possible "trajectory of development" of succession system and a few trajectories could be not enough representative for reliable prognosis.

Therefore, the research of asymptotic behaviour of large imitative gap models seems to be impossible without their inclusion to the frameworks of an analytical theory. Such inclusion can be carried out on the basis of the following thesis:

*succession gap-models can be considered as computer realizations of the structural models of successions.*

The synthesis of these two main types of succession models offers the actual problem, and a number of researchers have simultaneously approached the solution of this problem from the various viewpoints.

A simulation model using linkage between these two approaches was constructed in (Acevedo, Urban, Shugart 1996). Mathematical peculiarities of this model involve using the semi-Markovian process for the description of replacement process of one gap by another, which is caused by the significant spatial sizes of "elementary succession unit" in this model and, therefore, by necessity to describe time-dependent transient processes.

An alternative approach is based on the concept (Razumovsky 1981) that «an elementary unit of vegetation» is the area occupied by a single succession stage; such units having the area less than 1 m$^2$ were observed. As the area of «an elementary succession unit» could be rather small, the duration of a transient process of such units may also be small. This assumption allows us to construct a succession model within the framework of differential structural models of communities.

This concept was used in the analytical approach to the synthesis problems mentioned above realized in (Karev 1994) by the example of a structural model of succession dynamics. This model allows the description (at a phenomenological level) of the natural processes of dynamics of succession of forest communities within a given area, together with, and in relation to, internal dynamics. Specific cases are represented by Markov models and gap-models in the form of analytical structural models.

The values to be defined in a structural succession model are the areas, occupied at each time moment by a forest of certain «types», age and state. The concept of forest type includes species composition, site class, etc. Succession age means the time since the appearance of the forest in a given area. The forest state can be defined by current values of basic stand parameters (e.g., height and diameter, etc., of trees of a given age and species, and the stand density). The model of state dynamics is considered to be given and it is assumed that both the rate of area decrease (caused by self-thinning, fire, pests, felling, etc.) and the stochastic succession matrix of transition, specifying the succession model, are given. Elements of the succession matrix of transition are equal to the probabilities that a unit of area vacated by a type $j$ forest will be replaced by a type $i$ forest.

For the structural models of this kind, it has been shown that the natural dynamics of the community lead to a unique stable state (i.e., distribution of the areas, occupied by forests of defined types, age and status), which is attained at an exponential rate (Karev 1996).

Integral characteristics of forest cover (total area occupied by a forest of defined type, mean height and mean diameter of tree, wood stock of the area, etc.) are often of primary interest. In terms of structural model of succession all these characteristics can be written as some mean values of the area distribution, names by «regulators». Note that virtually all-essential characteristics of different ecosystems may be written in this way as the appropriate regulator.

It follows from the last results that the natural dynamics of the forest community lead to a unique value of any regulator at the stable state, which is attained at an exponential rate and can be calculated exactly.

So, in the stable state there are different processes of growth, mortality, change and renewal of vegetation cover on each fixed site, but the distribution of areas and other main characteristics of the forest ecosystem remains static.

The model is considered in more details below in s. 4, see also Mathematical Appendix.

## 2.9. Ergodic Properties of Steady State Forest Communities

One of the most interesting and useful approaches to the problems discussed in the previous section is based on the ideas of ergodic theory.

The "ergodic properties" of systems are well studied in physics owing to their numerous applications. The possibility of adaptations of ideas based on ergodic theory for biological problems has long been the subject of discussions. This would be especially important for forest ecology when studying the succession processes, the duration of which is normally much longer than the human life span.

According to ergodic ideas the climax (more exactly, stationary) state of succession system represents a spatially developed time history of succession association. This approach in different forms was known for a long time; as the "ergodic hypothesis in ecology" it was formulated in (Molchanov 1992): the areas $\Theta_i$ of the biocenoses making up a plant association should be proportional in a climax state of association to the proper times $T_i$ of their development in the succession series. Thus one obtain

$\Theta_i / T_i = K = const$ for all $i = 1,...n.$

(More precisely, the *proper time* $T_i$ represents the average time of renewal of the *i*-th biocenose in the climax state, see s. 4 below.)

The importance of the ergodic hypothesis is that the *area* distribution of the instantaneous process state estimating the current state of the process being developed in *time*. This is especially useful for studying the forest succession processes because the estimation of a forest ecosystem status may present a difficult problem, but the measurements made over the forest areas throughout a given year are quite possible with the help of an aerial photography.

Suppose that the ergodic hypothesis is valid and the *proper times* $T_i$ are known or can be calculated. Then it allows quite simple estimation of the deviation of the observed state of forest-ecosystem from the theoretically stable state. The areas and *proper times* of the biocenoses are put on the coordinate axes. Then the ergodic hypothesis asserts that in a climax state of association the points $(T_i, \Theta_i)$ will lie on a straight line passed through the origin. Hence, it becomes possible to estimate the difference between the observed and climax states of the system using the spread of these points from the line. All we need are: (1) to establish the ergodic hypothesis; (2) to find the "ergodic constant" which is found as a slope of the line, and (3) to calculate all necessary *proper times*. These problems have been solved in (Karev 1994, 1997).

Theoretical consideration of the validity of ergodic hypothesis and some other results outlined above should be done within the framework of a sufficiently generalized mathematical model, which admits one to strictly define and calculate all the necessary concepts and values of variables.

An appropriate choice of this model presents a very important item. The most relevant model, in the author's opinion, is a structural succession model (see s.4), which allows describing succession series (system) and its climax state as a stable state of the model as well as calculating the areas occupied by biocenoses at any given instant, characteristic times of biocenosis development, etc. The succession model has a hierarchical structure and includes as separate blocks the models of phytocenoses dynamics which may be very detailed (well-known gap-models serve as a good example).

General Ergodic theorem proved for structural succession model allows one to describe theoretically its steady state in detail. More precisely, the values of all "generalized variables" (which appears as some area distribution averages, for example, area distribution, average height, diameter, stock volume, etc. can be calculated for the model steady state.

In s. 4 we demonstrate the wide scope of above-outlined theoretical approach. The steady state of a succession system will be described on the basis of DMTG model of tree stand dynamics. Using this approach, one can apply the *ergodic method* to estimation of the deviation of the observed forest ecosystem state from the theoretically found stationary state thus elucidating the characteristic (having the form of "generalized variable") which gives rise to maximum deviation.

## 3. ANALYTICAL GROWTH MODELS

As was explained in s.2, the description of the steady state of a succession system given by the succession model proceed from the dynamic model governing forest conditions of different types.

There are a large number of different models of tree stands (see, for example, Shugart 1986; Botkin 1993; Korzukhin and Semevsky 1992; Antonovsky et al. 1991; etc; one of the latest review is given by A.Porte and H. H. Bartelink, 2002). Nevertheless, a problem of constructing appropriate simple dynamic models depending on a few interpretable parameters for describing only the main factors of the environment and giving as a result the basic forest taxonomic characteristics remains actual.

The model named DMTG (Karev, Skomorovsky 1999) offers a possible solution, which could be used in the hierarchical system of forest models. The main variables of the model are: (1) $N$, number of trees; (2) $H$, average height; (3) $D$, average diameter; (4) $w$, average volume of a single tree; (5) $W$, wood store, and 6) $O$, wood decay.

A model of tree number dynamics is particularly important for exact calculation of the wood store being grounded on the number of trees and the average volume of stems. Although there is many partially empirical models of tree stand numbers, the satisfactory theory of tree stand number dynamic is not found. Let us describe shortly a new approach to tree number dynamics modelling.

## 3.1. Inhomogeneous Dynamics of Tree Number: Models with Distributed Mortality Rate

The problem of dynamics of the tree number in a forest is one of the oldest and most important problems in forest ecology. Nevertheless, so far no satisfactory quantitative theory of the self-thinning process of tree stands has been developed. Number different processes of tree interactions and various environment conditions affect the processes of growth and death of trees in a complex way.

Even the problem of simulation of self-thinning of single breed even-aged tree stands, despite of major number of offered models, demands further developments. There exist two main types of relationships for even-aged tree population. The relationships of the first type link mean tree size and tree density to density-dependent mortality; the self-thinning formulas of the second type describe tree density as a function of tree stand age.

The most known example of the first type relationships is the "3/2 power law" and its modifications (Reineke 1933; Yoda et al. 1963; White 1981; Weller 1987; Zeide 1987, 1995; Lonsdale 1990; Tang et al. 1994; Ogawa 2001). This kind of relationships was used in many forest gap-models (Botkin 1993; Shugart 2004) and other individual-based models of plant populations.

A more sophisticated approach is based on modelling of local interactions between trees. The theoretical background is a concept of ecological field (Uranov1965; Mou et al., 1993; Li et al., 2000). Explanations of self-thinning in plant population have focused on plant shape and packing (Adler 1995), such that proposed dynamical model could be approximated to identify key parameters and relationships. Other simulation used indexes of competition for recourses, i.e. area, light, nutrition (Pacala, Silander 1985; Galitsky, Komarov 1987; Post, Pastor 1990; Ford, Sorrensen 1991; Huston, DeAngelis 1994, Zou, Wu 1995; etc.).

A number of tree interactions and various environment conditions affect the growth and death of trees in complex ways. Establishing the role of environmental stresses in tree mortality is difficult because of the multitude of interacting stresses affecting trees (Pedersen 1998; Franklin et al., 1987; Bossel, 1986; Manion, 1981). Variations in genetic structure also affect the mortality rate of trees (simulations were performed in Takahashi et al. 2000). Thus, it seems to be impossible to take into account all impacts on the death rate of trees in framework of a unit model.

Self-thinning formulas of the second type that describe the dependence of tree stand numbers on tree population age, are closer to the growth tables used in forestry (Zagreev et al. 1992). These kinds of formulas are also used in some forest gap-models (i.e., Prentice and Leemans 1990). Consider three equations that fit empirical data well:

1) Hilmi formula

$N(t) = N(0) \exp(-\alpha_0(1-\exp(-ct)))$,

$\alpha_0, c$ are constants.

2) Kayanus formula

$N(t) = (a+bt)/t^2$,

$a, b$ are constants.

3) Formulas of the "power" type

$N = ct^{-k}$

such as formulas of Terskov- Schmalgausent, Voropanov, etc., where $c, k$ are constants.

In all these formulas $t$ may be interpreted as time or as age of tree stand, as they describe the dynamics of even-age tree stands. These and other formulas together with references can be founded in survey (Terskov, Terskova 1980).

The mainstream in modelling of self-thinning is (explicit or implicit) assumption that the population of trees is uniform that is all trees are identical in the population. Apparently, it is difficult to expect further successes in simulation of tree number dynamics without waiving this wide spread, but excessively simplifying assumption that is both restrictive and unrealistic.

We'll consider a tree stand as a population of non-identical trees having different mortality rates dependently on the conditions and local environment. Let us note that this approach was considered and realized in some analytical and mainly computer simulating models (see, e.g., Adler 1996, Li et al. 2000; etc).

Our analytical approach is based on application of the general theory of inhomogeneous populations (Karev 2000) to the problem of self- thinning of tree stands (see details in (Karev 2002)). The main assumption is that *every tree has its own survival ability and mortality rate*. More generally, we'll consider population models with distributed values of the mortality rate $a$, such that every individual possesses its own *constant value* of the parameter $a$. The value of the parameter describes individual invariable property (such as vitality, a hereditary attribute or a specificity of the local habitat, which is constant during an individual life). The parameter remains unchanged for any given individual and may vary from one individual to another. Individuals in such population are not identical, differing in their parameter values, which makes the population not uniform.

In the remainder, it is shown that some known quite different formulas of forest stand self-thinning can be considered as solutions of inhomogeneous Malthus extinction model with corresponding initial distribution of the mortality rate of trees. Developed theory opens the way to improve them and to derive a lot of different self-thinning formulas using appropriate initial distributions of the mortality rate of trees.

More precisely, let us consider here the simplest case of well-known Malthusian model of extinction $dN/dt = -aN$, where $a>0$ is a mortality rate (Malthusian parameter).

Sometimes it is useful to study formally more general model of the form

$$dN/dt = -acN$$

where $c=const >0$ is a time-scaling parameter. The last equation is reduced to the standard Malthus model by the change of time: $t' = ct$.

Let us suppose now, that each individual of the population has its own extinction rate $a$ which is distributed over individuals, and its distribution in the initial instant is $P_0(a)$. Collecting together all individuals having the same value of "individual Malthusian parameter" $a$, we get the $a$-group; then let $l(t,a)$ denote the size of $a$-group at some moment $t$. Thus the total population consists of a set of disjoint $a$-groups of sizes $l(t,a)$.

The dynamics of such a population is described by the system

$$\mathrm{d}l(t,a)\,/\mathrm{d}t = -\ a\ l(t,a),$$
$$N(t) = \int_A l(t,a)\mathrm{d}a\ ,$$
$$l(0, a) = l_0(a), \tag{3.1}$$

where $A$ is the range of $a$. The initial distribution $l(0,a) = l_0(a)$ of the population parameter is assumed to be pre-specified.

For a given initial distribution of the Malthusian parameter, let us denote

$$L_0(t) = \int_A \exp(-at)\ P_0(a)\mathrm{d}a$$

(one could recognize in $L_0(t)$ the Laplace transform of $P_0(a)$). It was proved (Karev 2000) that for inhomogeneous population model (3.1) the next assertions are valid:

*the current population size $N(t)$ is determined by formula*

$$N(t) = N_0 L_0(t) \tag{3.2}$$

*and solves the equation*

$$\mathrm{d}N/\mathrm{d}t = -\ E_t a\ N, \tag{3.3}$$

*where $E_t a$ is the current mean of the Malthusian parameter;*

*(ii) the current parameter mean $E_t a$ satisfies the equation*

$$\mathrm{d}E_t a\ /\mathrm{d}t = -\sigma^2(t), \tag{3.4}$$

*where $\sigma^2(t)$ is the current parameter variance.*

It implies that modelling of inhomogeneous population dynamics on the base of the only average value of reproduction rate without taking into account it's distribution or at least it's variance may be essentially incorrect. Indeed, the dynamics of inhomogeneous populations with the same initial average of extinction rate can be absolutely different depending on initial distribution at whole. Let us point out that all real tree populations are inhomogeneous.

We have got also the description the evolution with time of the initial parameter distribution and its main characteristics, such as mean and variance. The mortality rate of inhomogeneous population is proportional to $E_t a$, the mean value of the Malthusian parameter at moment $t$ (see Eq.(3.3)). On the other hand, Eq. (3.4) shows that $E_t a$ decreases with a rate proportional to the current variance of $a$. This implies that the mortality rate of inhomogeneous population with any initial distribution decreases with time and becomes less then the mortality rate of corresponding standard Malthus model. The current variance and the current distribution at whole can be also computed.

Many important known probability distributions (and some non-standard that are of a special interest for purposes of self-thinning modelling) could be considered as initial ones and some new tree number formulas could be obtained.

Let us consider some known formulas of tree self-thinning from the point of view of the developed theory. The first step is to find such initial parameter distribution that considered empirical formula coincides with the solution of corresponding inhomogeneous extinction model. Actually it is an inverse problem of inhomogeneous population modelling: given the function $N(t)$, find a probability distribution $P(a)$ such that $N(t)$ is a solution of inhomogeneous model (3.1) with $P(a)$ as the initial distribution of the parameter. This problem has a simple solution given by the following statement.

Let the function $N(t)$ solves the inhomogeneous model (3.1) with initial distribution of the parameter $P_0(a)$. Then $P_0(a)$ is defined by formula

$$P_0(a) = L^{-1}[\ N(t)/\ N(0)],$$

where $L^{-1}$ is the converse Laplace transform.

Using the proposition, we can reveal the "hidden" initial distribution of the extinction rate for given $N(t)$. Then we can construct more appropriate $P_0(a)$ and to refine $N(t)$. Let us apply this approach to some known formulas for $N(t)$ (see proves and details in Karev 2000, 2002).

### *Hilmi Formula and its Refinement*

Let us suppose that the parameter $a$ in model (3.1) has the Poisson distribution with average $a_0$ in an initial instant. In such a case the distribution of the parameter at any moment of time is also Poisson with the mean $E_t a = a_0 \exp(-ct)$; the current population size is determined by the formula

$$N(t) = N_0 \exp(a_0(e^{-ct} - 1)), \tag{3.5}$$

which is exactly the Hilmi formula.

Thus, the Hilmi formula describes the Malthusian process of decreasing of the inhomogeneous population size; the population is divided into countable number of disjoint groups such that individuals from $i$-th group, $i=1,2,\ldots$ have the mortality rate $ic$ and the initial size of the group is

$$l_0(i) = N(0)\exp(-\alpha_0)(\ \alpha_0)^i/i!$$

It is difficult to give an acceptable interpretation of this model with infinite number of groups. Individuals from groups with large $i$ have unrealistic large extinction rate $ic$ and will be eliminated rapidly from the population; more of that, the initial size of such a group, $l_0(i) = N(0)\exp(-\alpha_0)(\alpha_0)^i/i!$ is less then 1 with large $i$. Thus, it is reasonable to consider the subdivision of a tree population only to finite number of groups that corresponds to the truncated Poisson distribution.

Let parameter $a$ can assume only finite number of values $i=0,1,\ldots k$. Then for all $t$ the distribution of the parameter $a$ is again the $k$-truncated Poisson distribution and the population size is

$$N(t) = N(0)C_0(k)\sum_{i=0}^{k}(\alpha_0 \exp(-ct))^i/i! \tag{3.6}$$

This formula is biologically more meaningful then initial Hilmi formula with $k = \infty$; it allows to fit real data more precisely. More of that, it is possible now to choose the unknown number of different $a$-groups $k$ by best fitting, for different $k$, to the observed time series of the population sizes. Thus we can estimate the number of groups of trees having different "surviving levels". For examples, computations for "normal" pine tree stands give the estimation of $k$=7-10. The accuracy of data fitness is better then 5% of mean-square deviation $S$. Fig.3.1 shows the model solution and real data for normal pine planting of 1st class of quality (Zagreev et al. 1992, Tables 130).

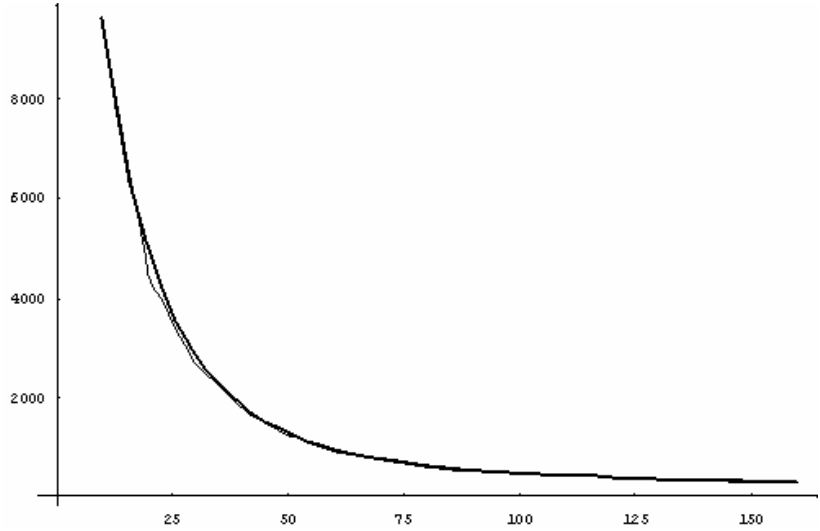

**Fig.3.1.** Data for pine planting of $1^{st}$ class of quality (thin) and modified Hilmi formula (3.1) (thick) with $k$=9, $\alpha_0$ =3.57, $c$=0.02, $S$=4.2%

### *The Kayanus formula:*

$$N(t) = (b_1 + b_2 t)/t^2 .$$
(3.7)

One can show that formally the Kayanus formula describes a Malthusian dynamics of decreasing inhomogeneous population where formally $l_0(a) = b_1 a + b_2$. The problem is that the linear function $l_0(a)$ can not be normalized to become lawful probability distribution. To overcome this difficulty, let us take into consideration that too large values of mortality rates are unrealistic. Thus it is reasonable to assume that $a \in [0,v]$, so that we can take

$$P_0(a) = C(v) \ (b_1 a + b_2), \ 0 < a < v$$
(3.8)

where normalizing constant $C(v) = (b_2 v + b_1 v^2/2)^{-1}$. It means that the initial distribution for modified Kayanus model has a linear pdf in $[0, v]$.

Then the following solution of inhomogeneous Malthus model corresponds to the distribution (3.8):

$$N(t)=N(0)\{ \ b_1 + b_2 t - \exp(-tv)(b_2 t + b_1 \ (1 + tv)\}/(t^2 \ v \ (b_2 + b_1 v/2))$$
(3.9)

The dynamics of tree stand number is described by the modified formula (3.9) much more precisely than by initial Kayanus formula, see Fig-s.2a, 2b. For example, the mean-square relative deviation of the settlement and table data on the dynamics of even-age normal pine tree stands of different classes of quality (Zagreev et al. 1992, Tables 130) does not surpass 4 %.

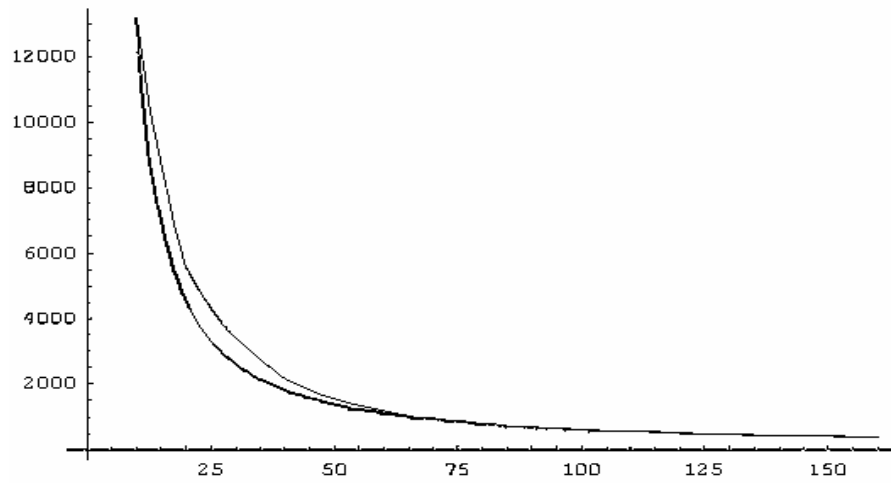

**Fig.3.2a.** Data for pine planting of 2$^{nd}$ class of quality (thin) and original Kayanus formula (3.2) (thick); $a$=18.1, $b$=1.1, $S$=9.5%

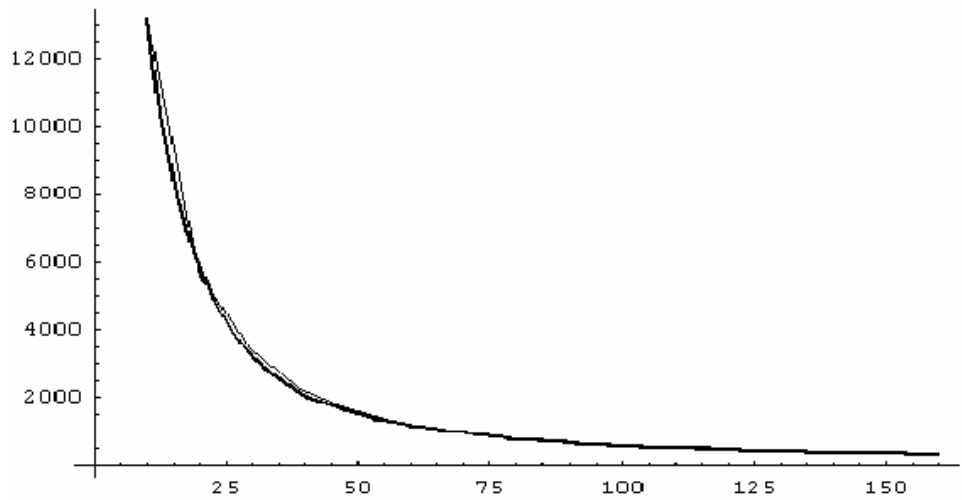

**Fig.3.2b.** Data for pine planting of 2$^{nd}$ class of quality (thin) and modified Kayanus formula (3.8) (thick); $v$=0.19, $a$=29.9, $b$=0.75, $S$=2.8%

Evolution of the initial linear distribution with time due to the modified Kayanus inhomogeneous model is shown on Fig.3.

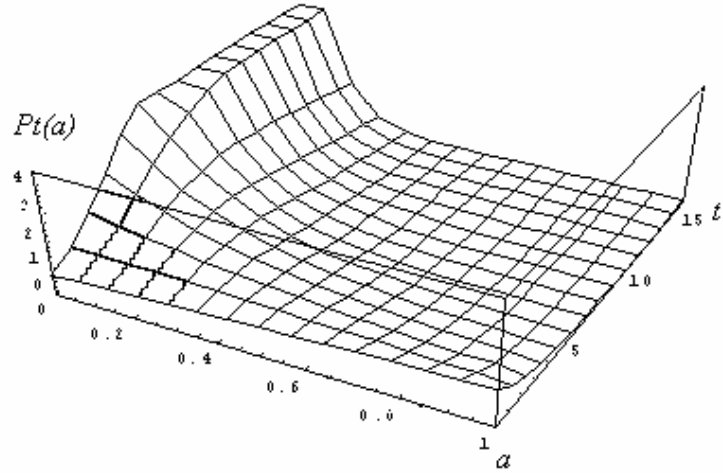

**Fig. 3.3**. Evolution of the linear initial distribution (3.7) with time according to formula (3.9) with $\nu$=0.19, $a$=29.9, $b$=0.75

### *The "Power" Formula of Schmalgausent-Terskov*

A typical data on self-thinning of pine tree stand in double logarithmic scale (Terskov, Terskova 1980) is close to a straight line; this graph corresponds well to the power dependence

$$N(t) = C\,(t + t_0)^{-k},\tag{3.10}$$

where $t_0$ is a starting time moment (or initial age), $C = N(0)t_0{}^k$.

Formula (3.10) describes Malthusian dynamics of decreasing inhomogeneous population where formally $l_0(a) \sim a^{k-1}$; again $l_0(a)$ can not be normalized to become lawful probability distribution. Thus suppose that $a \in [0,\nu]$ and consider the $k$-power distribution *in a bounded interval* $[0,\nu]$

$$P_0(a) = C(\nu)\,\alpha^{k-1},\ 0 < a < \nu$$

where normalizing constant $C(\nu) = k/\nu^k$. Then we get the refined formula

$$N(t) = N(0)\,k/\nu^k\,G(k,t)\,t^{-k}\tag{3.11}$$

where $G(k,t) = \displaystyle\int\limits_0^{t\nu} \exp(-s)s^{k-1}\mathrm{d}s.$

The dynamics of tree stand number can be described by the modified formula (3.11) more precisely, than by initial Schmalgausent-Terskov formula (3.10). For example, with $k=1.33$ and $v=0.27$ this formula fit the data of pine planting of $1^{st}$ class of quality (Zagreev et al. 1992, Tables 130) with mean-square deviation $S<4.5\%$.

The evolution with time of the "power" initial distribution due to the modified Schmalgausent-Terskov inhomogeneous model is shown on the Fig. 3.4.

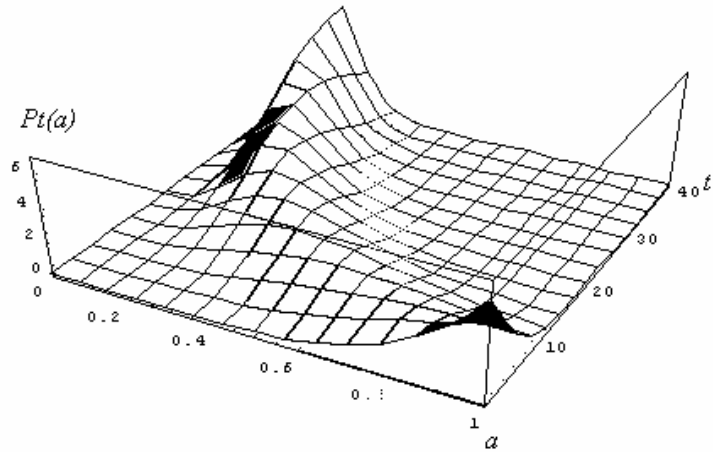

**Fig.3.4.** Evolution of the distribution (3.11) with $v=0.27$, $k=1.33$

## 3.2. Inhomogeneous Dynamics of Tree Number: Models with a Few Connected Groups

Let us consider another approach to analysis of the role of heterogeneity in tree populations. The approach is based on the division of the population into a few connected groups having different mortality rates. The division allows different interpretations for populations of trees such as: 1) a group corresponds to the age period (latent, pre-generative, generative and post-generative ones) or, more differentially, to the $k$-th "age condition" of trees; 2) a group corresponds to the "life level" (normal, lowered, low, sub-lethal, lethal). This approach was applied in (Karev, Skomorovsky 1997), (Karev, Berezovskaya 2002) to construct a self-thinning model of tree population based on Poletaev' approach (Poletaev 1980) to modelling of the "*Schmalgausent law*" (Schmalhgausent 1935).

The qualitative behaviour of the model accounts for the peculiarities of some experimental curves (such as the existence of "phases of development" and a point of inflection, see (Korzukhin, Semevsky 1992).

Suppose that a population can be shared on $m$ non overlapping groups with $n_1(t),...n_m(t)$ numbers and that survived individuals of the $k$-th group for $k<m$ transit to the $(k+1)$-th group with the intensity $s_k$. The next system of equations (a model of Leslie's type) follows from these postulates resulting in:

$$dn_1/dt = -s_1 n_1 - q_1 n_1,$$
$$dn_2/dt = s_1 n_1 - s_2 n_2 - q_1 n_2,$$

$\mathrm{d}n_{m-1}/\mathrm{d}t = s_{m-2}\,n_{m-2} - s_{m-1}\,n_{m-1} - q_{m-1}\,n_{m-1}$ ,

$\mathrm{d}n_m/\mathrm{d}t = s_{m-1}\,n_{m-1} - q_m\,n_m$ ,                              (3.12)

where $q_k$ is a death rate of individuals belonging to the *k-th* group.

Computer experiments proved that the best variant for tree populations is model (3.12) under *m*=3; the final model then takes the form:

$N(a) = n_1(a) + n_2(a) + n_3(a)$,

$\mathrm{d}n_1/\mathrm{d}a = -p_1\,n_1 - p_2\,n_1$ ,

$\mathrm{d}n_2/\mathrm{d}a = p_2\,n_1 - p_1\,n_2 - p_2\,n_2$ ,

$\mathrm{d}n_3/\mathrm{d}a = p_2\,n_2 - p_3\,n_3$                              (3.13)

under initial conditions $n_1(a_0) = N(a_0)$, $n_2(a_0) = n_3(a_0) = 0$. Here *a* is a current age, $a_0$ is an initial age (usually $a_0$=0), $N(a)$ is a whole number of trees having the age *a,* and $p_i$ are the model parameters.

The solution of the system (3.13) is given by the next formula $N(a) = N(a_0)\varphi(a)$ where

$\varphi(a)=\exp[-(p_1+p_2)(a-a_0)]*$                              (3.14)

$\{1+p_2(a-a_0)+p_2^2(a-a_0)/(p_3-p_1-p_2)-p_2^2/(p_3-p_1-p_2)^2*[1-\exp((p_1+p_2-p_3)(a-a_0))]\}$.

Here parameter $p_1$ may be interpreted as a death rate in juvenile and middle groups, $p_3$ as a death rate in old group, and $p_2$ as a rate of transition from one group to another.

Solutions $N(a)$ of model (3.13) with different parameter values taken from the Table 3.1 are shown on Fig.3.5.

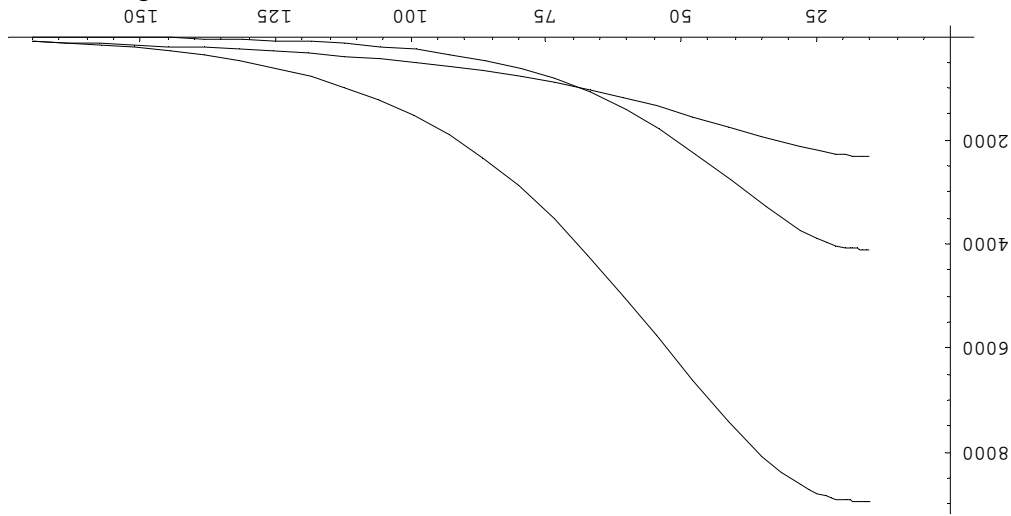

Fig.3.5. Solutions $N(a)$ of model (3.13) with parameter values taken from Table 3.1.

**Table.3.1. The parameter values of model (3.13); experimental areas of pine plantings on territory of a Forest Residence of the Timiryasev Academy, Russia (Advances 1964, table 24):**

| № | $a_0$ | $N(a_0)$ | $p_1$ | $p_2$ | $p_3$ | $S$ | $S_{max}$ |
|---|---|---|---|---|---|---|---|
| 1 | 15 | $2.31*10^3$ | $1.87*10^{-3}$ | $1.31*10^{-1}$ | $2.15*10^{-2}$ | 1 | 8 |
| 2 | 15 | $4.1*10^3$ | $5.64*10^{-4}$ | $1.03*10^{-1}$ | $4.66*10^{-2}$ | 2 | 14 |
| 3 | 15 | $8.95*10^3$ | $5.84*10^{-5}$ | $5.74*10^{-2}$ | $4.73*10^{-2}$ | 3 | 15 |

The values of parameters of model (3.13) for different breeds and classes of quality are given in the following Tables 3.2.- 3.4.

**Table 3.2. The parameter values of model (3.13) for normal pine planting (Zagreev et al. 1992, tables 130)**

| Class of quality | $N(a_0)$ | $p_1$ | $p_2$ | $p_3$ | $S$ | $S_{max}$ |
|---|---|---|---|---|---|---|
| 1б | 4390 | $4.28*10^{-2}$ | $2.22*10^{-2}$ | $6.86*10^{-3}$ | 1 | 13 |
| 1a | 7580 | $6.21*10^{-2}$ | $3.12*10^{-2}$ | $9.89*10^{-3}$ | 1 | 7 |
| 1 | 9650 | $6.51*10^{-2}$ | $3.09*10^{-2}$ | $1.01*10^{-2}$ | 1 | 14 |
| 2 | 11000 | $5.89*10^{-2}$ | $2.55*10^{-2}$ | $8.82*10^{-3}$ | 1 | 15 |
| 3 | 15600 | $6.43*10^{-2}$ | $2.78*10^{-2}$ | $1.01*10^{-2}$ | 1 | 11 |
| 4 | 9860 | $4.98*10^{-2}$ | $3.13*10^{-2}$ | $9.18*10^{-3}$ | 1 | 5 |
| 5 | 16000 | $5.18*10^{-2}$ | $2.61*10^{-2}$ | $8.05*10^{-3}$ | 1 | 13 |
| 5a | 12700 | $4.12*10^{-2}$ | $3.34*10^{-2}$ | $8.32*10^{-3}$ | 1 | 3 |
| 5б | 10300 | $2.71*10^{-2}$ | $3.33*10^{-2}$ | $5.83*10^{-3}$ | 1 | 2 |

**Table 3.3. The parameter values of model (3.13) for normal spruce planting (Zagreev et al. 1992, tables 131)**

| Class of quality | $N(a_0)$ | $p_1$ | $p_2$ | $p_3$ | $S$ | $S_{max}$ |
|---|---|---|---|---|---|---|
| 1б | 14600 | $7.01*10^{-2}$ | $2.30*10^{-2}$ | $1.09*10^{-2}$ | 1 | 10 |
| 1a | 14600 | $6.31*10^{-2}$ | $2.00*10^{-2}$ | $8.79*10^{-3}$ | 1 | 7 |
| 1 | 18000 | $6.50*10^{-2}$ | $2.07*10^{-2}$ | $8.88*10^{-3}$ | 1 | 7 |
| 2 | 24600 | $6.86*10^{-2}$ | $2.09*10^{-2}$ | $9.02*10^{-3}$ | 1 | 7 |
| 3 | 13500 | $6.29*10^{-2}$ | $3.26*10^{-2}$ | $8.95*10^{-3}$ | 2 | 9 |
| 4 | 15700 | $5.67*10^{-2}$ | $2.84*10^{-2}$ | $7.84*10^{-3}$ | 1 | 6 |
| 5 | 10600 | $4.53*10^{-2}$ | $3.32*10^{-2}$ | $6.71*10^{-3}$ | 1 | 5 |

**Table 3.4. The parameter values of model (3.13) for normal birch planting (Zagreev et al. 1992, table 133)**

| Class of quality | $N(a_0)$ | $p_1$ | $p_2$ | $p_3$ | $S$ | $S_{max}$ |
|---|---|---|---|---|---|---|
| 1a | $2.37*10^4$ | $1.3*10^{-1}$ | $3.09*10^{-2}$ | $1.12*10^{-2}$ | 1 | 10 |
| 1 | $3.5*10^4$ | $1.35*10^{-1}$ | $2.98*10^{-2}$ | $1.22*10^{-2}$ | 1 | 8 |
| 2 | $6.6*10^4$ | $1.51*10^{-1}$ | $2.7*10^{-2}$ | $1.32*10^{-2}$ | 1 | 10 |
| 3 | $9.43*10^4$ | $1.51*10^{-1}$ | $2.67*10^{-2}$ | $1.45*10^{-2}$ | 1 | 9 |
| 4 | $8.76*10^4$ | $1.22*10^{-1}$ | $2.48*10^{-2}$ | $1.35*10^{-2}$ | 1 | 10 |
| 5 | $14*10^4$ | $1.47*10^{-1}$ | $2.41*10^{-2}$ | $1.38*10^{-2}$ | 1 | 12 |

### 3.3. The DMTG Model

The basic variant of DMTG (Karev, Skomorovsky, 1999) has three independent variables: the number of trees $N$, the average height $H$ and the average tree volume $w$.

The model of tree number dynamics described in section 3.2 shows the best fitting of real data. So formula (3.14) makes up the first equation of DMTG. Remark, that the refined Kayianus formula (3.7) is almost of the same accuracy and could be used as the model of tree number dynamics.

A tree height is determined by the equation (Karev 1982):

$$dH/dt = h_1 E_2 - h_2 H^2 \tag{3.15}$$

where $E_2 = (1 - \exp(- h_3 N H^2))/(h_3 N H^2)$.

The basic equation of growth of a tree volume refers to an "energy balance" ascending to a work of Poletaev (1966):

$$dw/dt = v_1 E_1 - v_2 wH - v_4 B \tag{3.16}$$

where $E$, $B$ are the intensities of photosynthesis and breath, correspondingly; $v_2 wH$ is the charge on transport of assimilates; $v_1$, $vc_2$, $v_4$ are constants. It is reasonably safe to suppose that $B \sim E$, thus $v_4 = 0$.

The rate of photosynthesis $E_1$ can be taken proportional to the light energy intensity $J$ (in light-limited tree stands). Thus according to the model of Monsi and Saeki, one has

$$E_1 = (1 - \exp(- v_3 NL))/(v_3 N)$$

where $c_3$ is an extinction coefficient, $L$ is a photosynthetic tree surface. It is supposed that $L \cong H^d$ where $d$ is an estimated parameter; as a rule we put $d=2$. Thus, finally

$$dw/dt = v_1(1 - \exp(- v_3 NH^2))/(v_3 N) - v_2 wH \tag{3.17}$$

The model allows also calculating other important characteristics of tree stand, namely the average tree diameter $D$, the store of a tree stand $W$ and the size of total wood decay $O$. The Mitcherlich formula

$$H = b_1 + (b_2 - b_1)\{1 - exp[-b_3(D - D(t_0))/(b_2 - b_1)]\}$$

where $b_1$, $b_2$, $b_3$ are constants, is used for establishing the connection between height and diameter values:

$$D = D(t_0) + \ln[(b_2 - b_1)/ (b_2 - H)] [(b_2 - b_1)/b_3]. \tag{3.18}$$

The store $W(t)$ of a tree stand is calculated through an obvious formula

$$W = N\,w;\qquad(3.19)$$

the size of total wood decay $O(t)$ is determined by the equation

$$\mathrm{d}O/\mathrm{d}t = -\,ewdN/dt,\ e= const.\qquad(3.20)$$

Formulae (3.14)--(3.20) compose the complete DMTG system of equations.

Data on trial areas of Scots pine and Norway spruce stands (see Advances in Experimental Works… 1964), data on Scots pine plantations of various densities (Rubtsov et al. 1976, and also growth yield tables of "normal forest" (Zagreev et al. 1992) of various tree species (pine, spruce, larch, birch, oak, etc.), were used for verification of the DMTG. The mean-square deviation of the model from yield table data was less than 3-4 %. Almost the same accuracy was achieved when examining more complex variants of the model describing growth of two-species mixed stands.

As an example the following Table 3.5 presents the coefficients of the DMTG for normal Spruce plantings of different classes of quality (Zagreev et al.). The notions of $S$ and $S_{max}$ denote mean-square deviation and maximal deviation (in %) of corresponding computing values from data. The column "at age" indicates the age in which the maximal deviation happens.

**Table 3.5. The values of parameters of model DMTG**

| *Spruce, class of quality* **1b** | | | | | | | | |
|---|---|---|---|---|---|---|---|---|
| Number | | | Height | | | Store | | |
| $N(t_0)$ | $p_1$ | $p_2$ | $p_3$ | h1 | h2 | H3 | v1 | v2 | v3 |
| 14600 | 7.01e-2 | 2.30e-2 | 1.09e-2 | 5.08e-1 | 2.35e-4 | 2.30e-6 | 6.60e-1 | 7.27e-4 | 1.03e-4 |
| $S_{max}$ | at age | $S$ | | $S_{max}$ | at age | $S$ | $S_{max}$ | at age | $S$ |
| 10 | 150 | 1 | | 9 | 20 | 1 | 15 | 60 | 2 |
| *Spruce, class of quality* **1a** | | | | | | | | |
| Number | | | Height | | | Store | | |
| $N(t_0)$ | $p_1$ | $p_2$ | $p_3$ | h1 | h2 | H3 | v1 | v2 | v3 |
| 14600 | 6.31e-2 | 2.00e-2 | 8.79e-3 | 4.47e-1 | 2.52e-4 | 1.03e-5 | 6.03e-1 | 8.12e-4 | 9.53e-4 |
| $S_{max}$ | at age | $S$ | | $S_{max}$ | at age | $S$ | $S_{max}$ | at age | $S$ |
| 7 | 20 | 1 | | 9 | 20 | 1 | 17 | 30 | 3 |
| *Spruce, class of quality* **1** | | | | | | | | |
| Number | | | Height | | | Store | | |
| $N(t_0)$ | $p_1$ | $p_2$ | $p_3$ | h1 | h2 | h3 | v1 | v2 | v3 |
| 18000 | 6.50e-2 | 2.07e-2 | 8.88e-3 | 3.71e-1 | 2.45e-4 | 1.35e-7 | 4.80e-1 | 7.93e-4 | 9.75e-5 |
| $S_{max}$ | at age | $S\ (\%)$ | | $S_{max}$ | at age | $S\ (\%)$ | $S_{max}$ | at age | $S\ (\%)$ |
| 7 | 20 | 1 | | 12 | 20 | 1 | 17 | 70 | 2 |
| *Spruce, class of quality* **2** | | | | | | | | |
| Number | | | Height | | | Store | | |

| ***Spruce, class of quality 1b*** | | | | | | | | | |
|---|---|---|---|---|---|---|---|---|---|
| Number | | | | Height | | | Store | | |
| $N(t_0)$ | $p_1$ | $p_2$ | $p_3$ | h1 | h2 | H3 | v1 | v2 | v3 |
| 14600 | 7.01e-2 | 2.30e-2 | 1.09e-2 | 5.08e-1 | 2.35e-4 | 2.30e-6 | 6.60e-1 | 7.27e-4 | 1.03e-4 |
| $N(t_0)$ | $p_1$ | $p_2$ | $p_3$ | h1 | h2 | h3 | v1 | v2 | v3 |
| 24600 | 6.86e-2 | 2.09e-2 | 9.02e-3 | 3.14e-1 | 1.92e-4 | 4.03e-3 | 4.22e-1 | 4.37e-4 | 3.61e-3 |
| $S_{max}$ | at age | $S$ | | $S_{max}$ | at age | $S$ | $S_{max}$ | at age | $S$ |
| 7 | 20 | 1 | | 17 | 20 | 1 | 32 | 20 | 4 |
| ***Spruce, class of quality 3*** | | | | | | | | | |
| Number | | | | Height | | | Store | | |
| $N(t_0)$ | $p_1$ | $p_2$ | $p_3$ | h1 | h2 | h3 | v1 | v2 | v3 |
| 13500 | 6.29e-2 | 3.26e-2 | 8.95e-3 | 2.91e-1 | 3.57e-4 | 9.67e-9 | 3.33e-1 | 9.45e-4 | 5.15e-4 |
| $S_{max}$ | at age | $S$ | | $S_{max}$ | at age | $S$ | $S_{max}$ | at age | $S$ |
| 9 | 20 | 2 | | 8 | 30 | 1 | 9 | 20 | 1 |
| ***Spruce, class of quality 4*** | | | | | | | | | |
| Number | | | | Height | | | Store | | |
| $N(t_0)$ | $p_1$ | $p_2$ | $p_3$ | h1 | h2 | h3 | v1 | v2 | v3 |
| 15700 | 5.67e-2 | 2.84e-2 | 7.84e-3 | 2.10e-1 | 2.82e-4 | 2.26e-4 | 2.53e-1 | 8.47e-4 | 8.82e-5 |
| $S_{max}$ | at age | $S$ | | $S_{max}$ | at age | $S$ | $S_{max}$ | at age | $S$ |
| 6 | 40 | 1 | | 10 | 30 | 1 | 17 | 50 | 3 |
| ***Spruce, class of quality 5*** | | | | | | | | | |
| Number | | | | Height | | | Store | | |
| $N(t_0)$ | $p_1$ | $p_2$ | $p_3$ | h1 | h2 | h3 | v1 | v2 | v3 |
| 10600 | 4.53e-2 | 3.32e-2 | 6.71e-3 | 1.75e-1 | 4.04e-4 | 4.32e-4 | 2.18e-1 | 1.32e-3 | 1.98e-4 |
| $S_{max}$ | at age | $S$ | | $S_{max}$ | at age | $S$ | $S_{max}$ | at age | $S$ |
| 5 | 50 | 1 | | 5 | 40 | 1 | 13 | 40 | 1 |

# 4. SUCCESSION DYNAMICS AND THE STATIONARY STATE OF FOREST COMMUNITIES

## 4.1. Structural Model of Successions and its Equilibrium States

Meta-population approach, as it was described in 2.7-2.9 allows analytical modelling of forest dynamics in large space-temporal scales. There exist many different "landscape level" models and succession models (Horn 1975; Usher 1979; Binkley 1980; Shugart 2004; Shugart et al. 1988; Urban, Shugart 1992; Kellomäki et al. 1992; Kienast, Krauchi 1991; Leemans 1992; Levine et al. 1993; Korotkov et al. 2001; etc.). Let us underline that the succession models form an important branch of models intended for forest management (Mohren et al. 1991; Liu, Ashton 1995). In what follows we use the structural model of successions developed in (Karev 1994, 1996).

The quantities to be defined within a structural succession model are the areas $S_i(t,a,X)$ occupied at the instant $t$ by the forest of a certain type $i$, age $a$, and "condition" $X$. The notion of forest *type* includes species composition and classes of quality. Succession age $a$ defines the time elapsed from the moment of appearance of the forest over a given area (and generally $a$ is not equal to the age of tree stand, defined as average age of the main layer of the dominant species). The forest condition $X=(x_1,\ldots x_m)$ at the instant $t$ can be defined by current values of the basic stand parameters (such as average height, diameter, etc.) and a number of trees of a given age and species).

Let us note that the value $S_i(t,a,X)$ is a mathematical abstraction rather then a measurable value. Theoretically, it could be correlated with real data only if the "unit area" is very small (remark, that Razumovsky (1981) observed the "elementary succession units" having the area about 1 m$^2$). Practically, these values could be estimated only for conditionally even-aged forests. Nevertheless, the notion of $S_i(t,a,X)$ is useful for construction of a general mathematical model. The most reasonable interpretation is that $S_i(t,a,X)$ represents the number of (small) gaps or plots, which are occupied by a group (subpopulation) of trees having given values of variables of interest (age, height, diameter, volume, etc); the number of trees on the plots is also included in the "condition" $X$.

At this stage the dynamic model of forest condition is considered to be given. The known gap models may serve as the examples (Shugart 2004, Botkin 1993). Another example, the *DMTG* model, was described in Section 3.3 and is used below.

In the general case, the dynamics of the condition of the $i$-th type forest is described by the system of equations

$$\mathrm{d}X_i/\mathrm{d}a = F_i(a,X_i).$$

It is assumed that the decrease rates of the area (due to self-thinning, fires, pest impact, cuttings, etc.) are given. We suppose also that the stochastic matrix of succession transitions, specifying the succession model is known. Elements of the matrix of succession transitions are equal to the probabilities that a unit area vacated by the *j-th* forest type will be replaced with the forest of *i-th* type.

Let $v_k(a,X)$ be the rate of elimination (death) of the $k$-th type forest, $\gamma_{ik}(X)$ be the probability that a unit area released after the $k$-th type forest will be occupied by the forest of the *i-th* type, condition $X$ and zero age, and $\Omega=\{X\}$ be the set of all possible conditions.

The following model describes the dynamics of the forest succession system under consideration:

$$\mathrm{d}X_i/\mathrm{d}a = F_i(a,X_i),$$

$$\partial S_i/\partial t + \partial S_i/\partial a + \mathrm{div}(S_i\,F_i) = -v_i S_i,$$

$$S_i(t,0,X) = \sum_{k=1}^{n} \gamma_{ik}(X) \int\limits_{\Omega}\int\limits_{0}^{\infty} v_k(a,Y)\,S_k(t,a,Y)\mathrm{d}Y\mathrm{d}a,$$

$$\tag{4.1}$$

$$S_i(0,a,X) = S_i^0(0,X) \text{ is given.}$$

Assuming that the death rates $v_k$ and probabilities $\gamma_{ik}(X)$ of succession transitions are not explicitly time-dependent, the asymptotic behaviour of the system of equations (4.1) can be studied completely.

For the structural models of this kind it was shown, that the natural dynamics of the community leads to a unique stationary state, i.e. the distribution of the areas $S_i(a,X)$, occupied by forests of a definite type, age and condition, which is attained exponentially fast. The natural processes of growth, mortality, change and renewal of a vegetation cover proceed in the stationary state on each fixed plot, whereas the area distribution of the forest ecosystem as a whole remains stationary.

To formulate the main results, some additional notions are required. Let $X_i(a,Y)$ be the solution of the Cauchy problem

$$\mathrm{d}X_i/\mathrm{d}a = F_i(a,X_i),\ X_i(0) = Y,\ i=1,...n.$$

Then the expression

$$n_i(a,Y) = \exp\left(-\int_0^a v_i(u, X_i(u,Y))\,du\right)$$

is a survival function, i.e. the probability that a unit area of the $i$-type forest, whose initial condition is $Y$, will survive up to the age $a$.

Let us define the stochastic succession matrix

$$\Gamma_{ij} = \int_\Omega \gamma_{ij}(X)\mathrm{d}X$$

where $\Gamma_{ij}$ is the probability that a unit area released after the $j$-type forest will be occupied with the $i$-type forest, $i,j = 1,...n.$

We suppose that the succession matrix $\{\Gamma_{ij}\}$ is indecomposable.

Let $(u_1, ... u_n)$ be the right eigenvector (unique up to a constant factor) of the matrix $\{\Gamma_{ij}\}$, corresponding to the eigenvalue 1, $\sum_{j=1}^{n} \Gamma_{ij}u_j = u_i$; suppose additionally that this vector is normalized such that $\sum_{i=1}^{n} u_i = 1$. (As the matrix $\{\Gamma_{ij}\}$ is stochastic, $\sum_{i=1}^{n} \Gamma_{ij} = 1$, then its left eigenvector, corresponding to the eigenvalue 1, is the unit vector up to a constant factor).

Also define quantities:

$$\Theta_i(t) = \int_\Omega\int_0^\infty S_i(t,a,X)\mathrm{d}a\mathrm{d}X,$$

the total area of the $i$-th type forest at the instant $t$;

$$V = \sum_{k=1}^{n} \Theta_k(0),$$

the initial total area of the association;

$$m_i(X) = \int_0^\infty n_i(a,X)\mathrm{d}a,$$

the mean life span of a unit area of the *i*-th type forest with the initial condition $X$ (indeed, the mean life span of a unit area is, by definition,

$$m_i(X) = \int_0^\infty a\,v_i(a,X)n_i(a,X)\mathrm{d}a = -\int_0^\infty a(\mathrm{d}n_i(a,X)/\mathrm{d}a)\mathrm{d}a = \int_0^\infty n_i(a,X)\mathrm{d}a);$$

$$\tau_i(a,Y) = n_i(a,Y)\sum_k \gamma_{ik}(Y)u_k,$$

the survival function in the stationary state;

$$m = \sum_i \iint_{\Omega\ 0}^{\infty} \tau_i(a,Y)\mathrm{d}a\mathrm{d}Y = \sum_{i,j} \int_\Omega m_i(Y)\gamma_{ij}(Y)u_j\,dY,$$

the mean time of renewal of the succession system in the stationary state.

Assume that $m < \infty$, $V < \infty$ and for all $k$ and sufficiently large $a$

$$v_k(a,Y) > v = \text{const} > 0.$$

The studies (Karev 1994, 1996) succeeded in a complete description of the asymptotic behaviour of a succession system under the suppositions above: there exists only one asymptotically stable stationary state of system (4.1) that is attained exponentially from any initial state, which is described by a set of area distributions

$$S_i(a,X) = \lim_{t\to\infty} S_i(t,a,X),\ i=1,\dots n$$

(see Mathematical Appendix for the proof).

The ergodic hypothesis discussed in s. 2.9 is closely connected to that results and follow from them. Let us denote

$$\Theta_i = \iint_{0\ \Omega}^{\infty} S_i(a,X)\ \mathrm{d}a\mathrm{d}X$$

a total area of the *i*-th biocenosis in the stationary state, which is a value of interest. For computing this value, let us define the *proper time* of the *i*-th type forest:

$$T_i = \int\limits_{\Omega} \int\limits_{0}^{\infty} \tau_i(a,Y) \mathrm{d}a \mathrm{d}Y = \int\limits_{\Omega} [m_i(Y) \sum_{j=1}^{n} \gamma_{ij}(Y) u_j] \mathrm{d}Y,$$

which is equal to the average lifetime of a unit plot occupied by the *i*-th biocenosis in the stationary state of the system. The next theorem proves the ergodic hypothesis in the straightforward form.

**Ergodic theorem.** *Denote the ergodic constant by K = V/m. Then*

$$\Theta_i(t) \to \Theta_i \text{ with } t \to \infty, \text{ and } \Theta_i / T_i = K = V/m \text{ for all } i=1,...n. \tag{4.2}$$

## 4.2. General Ergodic Properties of the Stationary State of Forest Community

The ergodic hypothesis and more general assertion (4.2) prove to be true not only for the areas, but also for a wide class of the quantities, so-called *generalized variables*, which can be written in the form of some mean values computed over the area distribution:

$$Q_i(t) = \int\limits_{\Omega} \int\limits_{0}^{\infty} q(a,X) S_i(t,a,X) da dX \tag{4.3}$$

or

$$G_i(a,t) = \int\limits_{\Omega} q(a,X) S_i(t,a,X) dX .$$

Here $Q_i(t)$ is the generalized variable for the *i*-th biocenosis at the instant *t*. Any generalized variable is defined by the non-negative weight function $q(a,X)$. For example, if $q \equiv 1$, then $Q_i(t)$ is equal to the total area occupied by the *i*-th biocenosis at the instant *t*. Let us remark that theoretically all the relevant characteristics of the forest ecosystem, such as the average tree height and diameter; the areas occupied by trees of a given age; the growing stock, etc. could be expressed as appropriate generalized variables.

To formulate the general ergodic theorem, let us introduce the *proper times*:

$$[TQ_i] = \int\limits_{\Omega} \int\limits_{0}^{\infty} q(a,X_i(a,Y)) \tau_i(a,Y) da dY ,$$

$$[TG_i(a)] = \int\limits_{0}^{\infty} q(a,X_i(a,Y)) \tau_i(a,Y) dY .$$

$$\tag{4.4}$$

These quantities stand for the proper times of the generalized variables *Q* and *G,* correspondingly.

Assume that some known conditions providing the possibility of passage to the limit under the integral in (4.4) are fulfilled. Then the following theorem is valid

### General Ergodic Theorem

*For all generalized variables Q or G under conditions stated above there exist the limit values*

$$\lim_{t\to\infty} Q_i(t) = \mathbf{Q}_i = \int\limits_{\Omega} \int\limits_{0}^{\infty} q(a, X)\mathbf{S}_i(a, X)\,da\,dX,$$

$$\lim_{t\to\infty} G_i(a,t) = \mathbf{G}_i(a) = \int\limits_{\Omega} q(a, X)\mathbf{S}_i(a, X)\,dX$$

(4.5)

such that for every *i*

$$\mathbf{Q}_i/[TQ_i] = K,$$
$$\mathbf{G}_i(a)/[TG_i(a)] = K.$$

(4.6)

The theoretical importance of this ergodic theorem lies also in the reverse statement: from the validity of equation (4.6) for all generalized variables it follows that the succession system resides in the stationary (climax) state.

## 4.3. Ergodic Method of Complex Estimation of the Stationary State

*The ergodic theorem* asserts that all biocenosis characteristics (generalized variables) in the stationary state turn out to be proportional to their own proper times. In (4.6) $Q_i$ is the generalized variable in the stationary state of the forest association for *i*-th biocenosis that appears as a particular average taken over the limit area distribution $S_i(a,X)$, and $[TQ_i]$ is the proper time of the generalized variable for *i*-th biocenosis, which can be calculated exactly within the structural model of succession. The constant $K = V/m$ is *universal* for a given model and is equal to the common area of the succession system $V$, divided to $m$, the mean time of renewal of the succession system in the stationary state.

Equation (4.6) serves as the foundation of theoretically simple *method of complex estimation* of the deviation of the observed state of the forest ecosystem from the theoretical stationary state.

Let us lay off the values of any generalized variable $Q_i$ for all components of the succession system in its climax state on the *y*-axes and the values of corresponding proper times $[TQ_i]$ on the *x*- axes. Then the set of points ($[TQ_i]$, $Q_i$) on the plane belong to the common straight line for every generalized variable.

The quantity

$$D^2[Q(t)] = \sum_i \left( Q_i(t)/[TQ_i] - K \right)^2,$$

where $Q_i(t)$ are the observable values of a given characteristic for $i$-th biocenosis at the instant $t$ may be taken as the measure of a deviation (by characteristic $Q$) of an observable state of the forest community from the equilibrium climax state.

Hence, it is possible to find out which characteristic of the real forest community (e.g., stock of wood or age structure) deviates most strongly from a theoretical stationary state, and to estimate the degree of the deviation. This might be especially valuable for estimating the status of reserve lands and national parks and for computer processing data on ecological monitoring of natural forest ecosystems.

## 4.4. Distributions of the Main Taxonomic Values in a Stationary State of Forest Ecosystem

Consider the structural model of a forest ecosystem based on the DMTG model using the suppositions that forest community consists of $n$ types of forests; forest type is defined by main breed, which is characterized by density (tree number per unit area) $N$, average height $H$, and average tree volume $w$; equations (3.14), (3.15) and (3.17) describe dynamics of the main variables $N$, $w$ and $H$; the equation parameters depend on the forest type number.

For construction of the specific structural model of dynamic behaviour of forest territory, defining the distribution of areas $S_i(t,a,N,H,w)$, it is necessary to specify functions $\gamma_{ij}$ and $v_j$, which find application in the definition of the survival function $n_i(a,k)$, the probability of surviving of a plot of the unit area and the age $a$, where $k$ is a tree number at the initial moment. In considered models we suppose that the survival function of a single tree of $i$-type forest, $\varphi_i(a)$, depends only on the tree age, then the survival function of a plot is

$$n_i\,(a,k) = 1 - (1 - \varphi_i(a))^k.$$

Assume that the initial values $w_i^0$, $H_i^0$ are fixed for each $i = 1\ldots n$ and that the initial number of $i$-type trees is distributed according to the Poisson plain distribution with the mean $\lambda_i$. Thus

$$p_i\,(k) = \exp(-\lambda_i)\ \lambda_i^{\,k}/k!$$

stands for the probability that the initial number of trees per unit area is equal to $k$.

Applying the general results of s.4.1, let us calculate the ergodic constant for the model, as well as the mean values of the main variables specifying in the stationary state of the forest ecosystem and corresponding proper times.

The average life span of a unit patch of $j$-type forest with the initial number $k$ is written as

$$m_j\,(k) = \int\limits_0^\infty\ n_j\,(a,k)\,da = \int\limits_o^\infty\ [1 - (1 - \varphi_j(a))^k]\,da;$$

the survival function of a unit patch of $j$-type forest with the initial number $k$ in a stationary state reads as

$\tau_j\,(a,k) = u_j\,n_j\,(a,k)\,p_j(k) = u_j\,[1 - (1 - \varphi_j(a))^k]\exp(-\lambda_j)\,\lambda_j{}^k/k!.$

It is easy to compute that $\sum\limits_{k=0}^{\infty}\tau_j(a,k) = u_j(1 - \exp(-\lambda_j\varphi_j(a)))$. As $\lambda_j$ is the mean initial number of $j$-type trees, the value $\lambda_j\varphi_j(a)$ is equal to the number of trees of $j$-type forest at age $a$ (per unit area), i.e. $N_j(a) = \lambda_j\varphi_j(a)$ and so

$$\sum_{k=0}^{\infty}\tau_j\,(a,k) = u_j(1 - \exp(-N_j(a))).$$

Thus, the mean time of renewal of the forest system in the stationary state equals

$$m = \sum_{j=1}^{n}\int_{o}^{\infty}\sum_{k=0}^{\infty}\tau_j\,(a,k)\,\mathrm{d}a = \sum_{j=1}^{n}u_j\,m_j \tag{4.7}$$

where $m_j = \int\limits_{o}^{\infty}[1 - \exp(-N_j(a))]\mathrm{d}a.$

The universal *ergodic constant* for a given model is

$$K = V/m \tag{4.8}$$

where $V$ is the total area of the forest ecosystem.

The distribution of areas $\Theta_i$ occupied by $i$-type forest in the stationary state of the ecosystem is defined by equalities

$$\Theta_i = K\,T_i, \tag{4.9}$$

where according (4.4) (with $q=1$) the "proper times of areas" are

$$T_j = u_j\int_{o}^{\infty}[1 - \exp(-N_j(a))]\mathrm{d}a. \tag{4.10}$$

The *age distributions,* $S_j(a)$, where $S_j(a)$ is the area occupied by trees of the age $a$ of the $i$-type forest in the stationary state are given by equalities

$$S_j\,(a) = K\,[T_j\,(a)] = K\,u_j\,[1 - \exp(-N_j(a))],$$

as the "proper times of the age distribution" are

$$[T_j\,(a)] = u_j\,[1 - \exp(-N_j(a))].$$

From here, by substituting $K$ from (4.9), we obtain for any age

$$S_j(a)/\left[1-\exp(-N_j(a))\right] = \Theta_j/\int\limits_o^\infty \left[1-\exp(-N_j(a))\right]\mathrm{d}a \tag{4.11}$$

This equality shows the fraction of the total area of a *j*-type forest occupied by trees of the age *a*.

General formulas (4.4) for "proper times" read now

$$[TQ_j] = u_j \int\limits_\Omega q(a, X_j(a))[1 - \exp(-N_j(a))]\mathrm{d}a,$$

$$[TG_j(a)] = u_j\, g(a, X_j(a))[1 - \exp(-N_j(a))]. \tag{4.12}$$

Next, according to the Ergodic theorem (4.6), for any "generalized variable" $Q_j$ we have in the stationary state of the system

$$\boldsymbol{Q}_j/\left[\, u_j \int\limits_o^\infty q(a, X_j(a))(1 - \exp(-N_j(a)))\mathrm{d}a\right] = K,$$

$$\boldsymbol{G}_i(a)/\left[\, u_j\, q(a, X_j(a))(1 - \exp(-N_j(a)))\right] = K. \tag{4.13}$$

These formulas can be rewritten also as

$$\boldsymbol{Q}_j = \int\limits_o^\infty q(a, X_j(a))\, S_j(a)\mathrm{d}a,$$

$$\boldsymbol{G}_i(a) = q(a, X_j(a))\, S_j(a). \tag{4.14}$$

Thus we have the following formulas completely describing (together with formulas (4.9) and (4.12)) the stationary state of the forest ecosystem.

The mean height of trees of the *j*-type forest in the stationary state is

$$\boldsymbol{H}_j = \int\limits_o^\infty \int\limits_\Omega H\, S_j(a,N,H,w)\, \mathrm{d}N\, \mathrm{d}H\, \mathrm{d}w\, \mathrm{d}a\, /\Theta_j,$$

where $S_j(a,N,H,w)$ is the area distribution of the j-type forest in a stationary state, $\Omega$ is the range of variables *N, H,* and *w.* According to (4.13)

$$\boldsymbol{H_j} \, \Theta_j \, / [ \, u_j \int\limits_o^\infty \, H_j(a)(1-\exp(-N_j(a))) \mathrm{d}a] = K \text{ , or}$$

$$\boldsymbol{H_j} = \int\limits_o^\infty \, H_j(a) \, \mathrm{S}_j(a) \mathrm{d}a \, / \Theta_j =$$

$$\int\limits_o^\infty \, H_j(a) \, [1-\exp(-N_j(a))] \, \mathrm{d}a \, / \int\limits_o^\infty \, [1-\exp(-N_j(a))] \mathrm{d}a \tag{4.15}$$

where $H_j(a)$ is the solution to equation (3.15).

The store of wood of the $j$-type forest in the stationary state of a forest ecosystem is

$$\boldsymbol{W_j} = \int\limits_o^\infty \int\limits_\Omega \, [Nw \, S_j(a,N,H,w) \, \mathrm{d}N \, \mathrm{d}H \, \mathrm{d}w] \, \mathrm{d}a.$$

So

$$\boldsymbol{W_j} / [ \, u_j \int\limits_o^\infty \, W_j(a)(1-\exp(-N_j(a))) \, \mathrm{d}a] = K \tag{4.16}$$

where $W_j(a) = N_j(a) \, w_j(a)$ and $w_j(a)$ is the solution to equation (3.17), and

$$\boldsymbol{W_j} = \int\limits_o^\infty \, W_j(a) \, \mathrm{S}_j(a) \mathrm{d}a.$$

The store of wood of a $j$-type forest in a given age $a$ is $\boldsymbol{W_j}(a) = W_j(a) \, \mathrm{S}_j(a)$, thus $\boldsymbol{W_j}(a) \, / [ \, u_j \, W_j(a)(1-\exp(-N_j(a)))] = K$.

Now we can see that all generalized variables in the stationary state of the system are closely connected by the following ergodic relations:

$$\Theta_j \, / \, [u_j \int\limits_o^\infty \, (1-\exp(-N_j(a))) \mathrm{d}a] =$$

$$\mathrm{S}_j(a) \, / \, [ \, u_j \, (1-\exp(-N_j(a)))] = \boldsymbol{H_j} \Theta_j \, / \, [u_j \int\limits_o^\infty \, H_j(a)(1-\exp(-N_j(a))) \mathrm{d}a] =$$

$$\boldsymbol{W_j} \, / \, [ \, u_j \int\limits_o^\infty \, W_j(a)(1-\exp(-N_j(a))) \, \mathrm{d}a] = \boldsymbol{W_j}(a) \, / \, [ \, u_j \, W_j(a)(1-\exp(-N_j(a)))] = K \tag{4.17}$$

where $K = V/m = V / \sum\limits_{j=1}^{n} u_j \, m_j$ is the ergodic constant.

**Remark**. Let us underline, that the real duration of succession stages is finite; it means that $n_i(a,Y)$, a survival function, $S_i(t,a,Y)$, the area distribution, etc. vanish out an interval $[a_{st},$ $a_{fin}]$ where $a_{st}$ and $a_{fin}$ are the initial and final age of corresponding succession stage. Hence in all formulas above integrals over $a$ should be taken not from 0 to $\infty$, but from $a_{st}$ to $a_{fin}$.

Let us note that the structural model of forest ecosystems, in which the dynamics of every forest type is described by the equations of any gap model (Botkin 1993, Shugart 2004), could be constructed similarly. As a result we could get an analytical version of the gap-model of a landscape level; a full description of its stationary state can be achieved in explicit analytical form similarly with the help of formulas (4.8), (4.9), (4.15), (4.16). The resulting model allows to define directly the limiting distributions and to find limiting values of "generalized variables", passing a stage of computer calculations of transitive regimes.

Moreover, it becomes possible to investigate the dependence of the asymptotic distributions of various parameters of the model, by other words, to analyse the climax states of forest communities in line with the various scenarios of climatic and anthropogenic factors.

## 4.5. Applications of the Ergodic Theorems

### 4.5.1. *A Simplified Algorithm*

The ergodic relations (4.17) completely describe the stationary state of the forest ecosystem under suppositions that the initial values of the main variables (height, volume, etc.) are fixed and the initial tree numbers are distributed according the Poisson distribution for all succession stages of the forest ecosystem.

A simplified algorithm for computation of characteristics of the stationary state of the succession system follows from these formulas.

At first, we need the succession matrix $\{\Gamma_{ij}\}$ and its left eigenvector $(u_1, \dots u_n)$ corresponding to the eigenvalue 1.

Secondly, we should compute the values $m_j = \int\limits_{a_{st}}^{a_{fin}} [1 - \exp(-N_j(a))] \mathrm{d}a$ where $a_{st}$ and $a_{fin}$ are the initial and final age of $j$-th succession stage. The values of tree number $N_j(a)$ should be calculated with the help of appropriate model of tree stand self-thinning, e.g., described in s. see s.5.3 or could be taken from a real data. For preliminary estimations we can simply use *Forest Growth Tables* as it is done below.

Third, we should calculate the fundamental ergodic constant $K = V/m$, where

$m = \sum\limits_{j=1}^{n} u_j \, m_j$ and $V$ is the total area of the forest ecosystem.

Finally, we apply formulas of the previous section to compute the characteristics of interest of the stationary state of a forest ecosystem. In particular, for computation of the stationary area distribution we should

(i).        to define "proper times" $T_i = u_i \, m_i$ for each stage $i = 1, \dots k$ ;

(ii).     to postpone the set of points $\{(T_i, KT_i)\}$ with $x$-coordinates $T_i$ and $y$-coordinates $KT_i$, $i=1,\ldots k$ on the "ergodic line" $y=Kx$ (with $x=T$, the proper time, and $y=S$, the area), which show the theoretical stationary values of areas of corresponding stages;

(iii).    to put on the plane $(x,y)$ the set of points $\{T_i, S_i(t)\}$, $i=1,\ldots k$ where $S_i(t)$ is the value of area of the $i$-th stage at time $t$ and estimate the deviation of this set of points from the ergodic line $y=Kx$.

Let us underline again that the theoretical stationary values of the stage areas should be proportional to the "proper times" $T_i=u_i m_i$ for each stage; in turns the proper time is factorised into $u_i$, the component of the eigenvector of the succession matrix, and $m_i$, the average life span of a unit plot of the $i$-th stage, which does not depend on the succession matrix and is calculated with the help of a model or *Forest Growth Tables*.

### 4.5.2. *Successions in Priosko-Terrasnyi Biosphere Reserve*

Let us consider as a demonstration example the mixed boreal forest located in Priosko-Terrasnyi Biosphere Reserve, Moscow region, Russia. The model of natural course of succession through forest types in this Biosphere Reserve was constructed in (Logofet, Korotkov 2002). Authors studied 15 stages of natural succession; the succession matrix was constructed by the usual stage-duration method (Horn 1975; Cherkashin 1981; etc.): the probability of transition out of a stage supposed to be inverse proportional to the stage duration. The last stage, polydominant spruce-broadleaf forest is considered in the model as a climax stage and corresponds to the "absorbing" state of the model. It means that in course of "the model succession" this uniform polydominant forest will occupy the total area.

On our opinion, the notion of absorbing climax stage is a theoretical or mathematical abstraction rather than real observed state. Such a state should be destroyed in course of time by forest fires and/or insect attacks on large areas or as a result of natural succession dynamics of small plots, which transit them to one of the previous succession stages. Real data shows that it is difficult to find a stationary climax stage of that type even in reserve lands (including the Priosko-Terrasnyi Biosphere Reserve). On the other hand, it is possible to consider the same polydominant forest not as a uniform climax forest cover but as a large set of small plots being on different stages of development – actually it depends on the choice of the space scale. Thus we prefer to follow the classic, Clementsian, paradigm of succession theory (which is of course a theoretical abstraction also) that considers the *limit distribution* of areas as the dynamical climax state of ideal course of succession. The successful prediction of the "final" distribution of areas essentially depends on the appropriate choice of the succession matrix.

Let us consider a natural course of succession of the Priosko-Terrasnyi Biosphere Reserve from this point of view. According data (Logofet, Korotkov 2002, Table 1), the last 5 succession stages considered in this paper did not occupy any area in 1955 and also in 1999 years, so we'll study a simplified scheme of natural course of succession (Korotkov, personal communication; see Fig. 4.1).

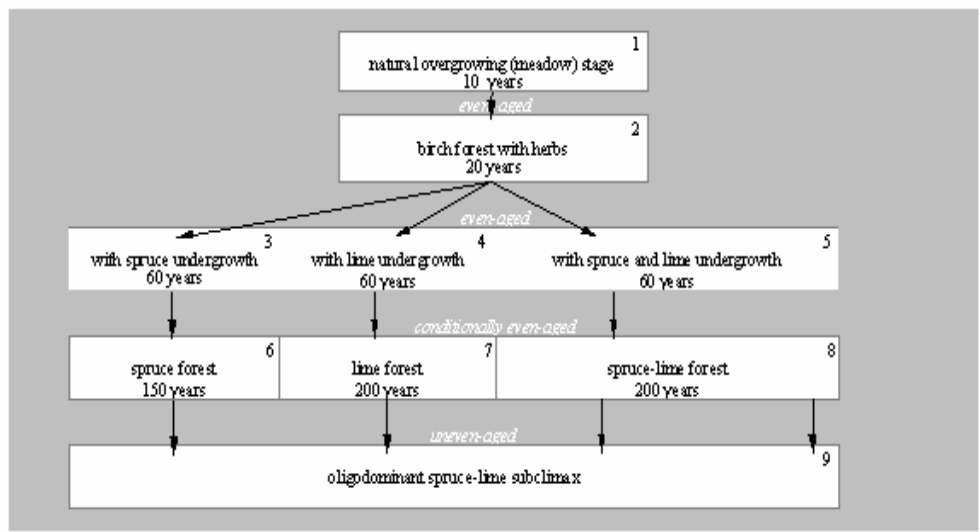

**Fig.4.1**. Simplified scheme of natural succession in the Prioksko-Terrasnyi Biosphere Reserve

The main breads are birch of the 1[st] and 2[nd] classes of quality, spruce of the 1[st] and 2[nd] classes of quality, lime of the 1[st] and 2[nd] classes of quality. The following Table presents the durations of each stage.

Below we will consider two versions of the course of natural succession given in the table 4.1. The first, *full* version takes into account 9 stages including the theoretical "climax" stage 9; the second, *short* version takes into account only the first 8 stages.

### Table 4.1. Data on simplified succession scheme in Prioksko-Terrasnyi Biosphere Reserve

| № of stage | Duration of stage, Years | Main breed | $a_{st}$ | $a_{fin}$ |
|---|---|---|---|---|
| 1 | 10 | Birch | 1 | 10 |
| 2 | 20 | Birch | 11 | 30 |
| 3 | 60 | Birch | 31 | 90 |
| 4 | 60 | Birch | 31 | 90 |
| 5 | 60 | Birch | 31 | 90 |
| 6 | 100 | Spruce | 60 | 160 |
| 7 | 140 | Lime | 60 | 200 |
| 8 | 100 | Spruce-Lime | 60 | 160 |
| 9 | | Spruce-Lime | 0 | 200 |

As the life duration of spruce and lime do not exceed 200 years, we could suggest that the last "climax" stage is also destroyed (at least partly) at this time. Further, we might suppose that the free area after the 2[nd] stage is distributed uniformly between the 3-5 stages but it could be not appropriate simplification. Actually we have no justified rule of this distribution; nevertheless the observed proportions of areas occupied by stages 3-5 could be considered as an approximate estimation of such a rule. Thus, let us suppose in what follows that the free area after the 2<sup>nd</sup> stage is distributed between the 3-5 stages proportionally to the areas of 3-5 stages in 1999, i.e. 26 : 34.2 : 37.4 (see Table 4.3).

Then, if to suppose that the free area transit to the first stage, we get the following succession matrix $\Gamma_1$:

$$\Gamma_1 = \begin{pmatrix}
0.9 & 0. & 0. & 0. & 0. & 0. & 0. & 0.01 & 0.005 \\
0.1 & 0.95 & 0. & 0. & 0. & 0. & 0. & 0. & 0. \\
0. & 0.014 & 0.983 & 0. & 0. & 0. & 0. & 0. & 0. \\
0. & 0.019 & 0. & 0.983 & 0. & 0. & 0. & 0. & 0. \\
0. & 0.017 & 0. & 0. & 0.983 & 0. & 0. & 0. & 0. \\
0. & 0. & 0.017 & 0. & 0. & 0.99 & 0. & 0. & 0. \\
0. & 0. & 0. & 0.017 & 0. & 0. & 0.993 & 0. & 0. \\
0. & 0. & 0. & 0. & 0.017 & 0. & 0. & 0.99 & 0. \\
0. & 0. & 0. & 0. & 0. & 0.01 & 0.007 & 0. & 0.995
\end{pmatrix}$$

The eigenvector of this matrix is $\boldsymbol{U_1}=(u_1, \ldots\ldots u_9) = (0.03, 0.06, 0.051, 0.067, 0.061, 0.08, 0.16, 0.1, 0.39)$.

By the reasons explained above let us exclude the last "climax" stage, polydominant forest, from the course of successions. The following succession matrix takes into account stages 1-8, see Fig.4.1 and Table 4.1; note that the "climax" stage 9 is absent in the real data in Tables 4.2, 4.3. Instead, we suppose that area vacated after 6-8 stages is occupied by the first stage. Then we get the following succession matrix $\Gamma_2$:

$$\Gamma_2 = \begin{pmatrix}
0.9 & 0. & 0. & 0. & 0. & 0.01 & 0.007 & 0.01 \\
0.1 & 0.95 & 0. & 0. & 0. & 0. & 0. & 0. \\
0. & 0.014 & 0.983 & 0. & 0. & 0. & 0. & 0. \\
0. & 0.019 & 0. & 0.983 & 0. & 0. & 0. & 0. \\
0. & 0.017 & 0. & 0. & 0.983 & 0. & 0. & 0. \\
0. & 0. & 0.017 & 0. & 0. & 0.99 & 0. & 0. \\
0. & 0. & 0. & 0.017 & 0. & 0. & 0.993 & 0. \\
0. & 0. & 0. & 0. & 0.017 & 0. & 0. & 0.99
\end{pmatrix}$$

The eigenvector of $\Gamma_2$ is $\boldsymbol{U_2}=(u_1, \ldots u_8)=(0.049, 0.98, 0.083, 0.11, 0.1, 0.14, 0.255, 0.017)$.

We can estimate the time, which is necessary to achieve the state close to the theoretical stationary state. A simplest way is based on calculation of the $n$-th power of the succession matrix; then every raw (of the matrixes $P$) tends to the desirable eigenvector. It is easy to see by this way that after 300 years the percent of each area stage is achieved with accuracy up to 0.5-0.7%; after 400 years the stationary area distribution is achieved with accuracy up to 0.1%. Thus the characteristic time for stabilizing the forest system is about 350-400 years.

Next, let us calculate the ergodic constant according formula $K = V/m = V/\sum_{j=1}^{n} u_j\, m_j.$

The integral $m_j = \int\limits_{a_{st}}^{a_{fin}} [1-\exp(-N_j(a))]\mathrm{d}a$ could be calculated with the help of the model of

tree number dynamics (see s.3) or estimated numerically (by plain trapezium formula):

$$\int\limits_{a_{st}}^{a_{fin}} [1-\exp(-N_j(a))]\mathrm{d}a \cong \Delta\{1/2([1-\exp(-N_j(a_{st}))]+[1-\exp(-N_j(a_{fin}))])+ \sum\limits_{a_{st}+\Delta}^{a_{fin}-\Delta} [1-\exp(-N_j(a_k))]\}$$

Here $\Delta$ is an age interval (10 or 5 years), $N_j(a_k)$ is the number of trees (per unit area) of the $j$-th breed at $a_k$ age and $a_k$ change from the initial age $a_{st}$ of the main breed in the stage to the final age $a_{fin}$, see Table 4.1. For preliminary estimation, we follow the second way and use the *Forest Growth Tables,* (Zagreev et al.1992).

Thus we get the following estimations of $m_j$:

*stage* 1), birch of the 1$^{st}$ class of quality, $\Delta = 5$, $a_{st}=1$, $a_{fin}=10$:
$m_1=15.5$;

*stage* 2), birch of the 1$^{st}$ class of quality, $\Delta = 5$, $a_{st}=11$, $a_{fin}=30$:
$m_2=11.82$;

*stages* 3-5), birch of the 1$^{st}$ class of quality, $\Delta = 5$, $a_{st}=31$, $a_{fin}=90$:
$m_3=m_4=m_5=11.23$;

*stage* 6), spruce of the 1$^{st}$ class of quality, $\Delta = 10$, $a_{st}=61$, $a_{fin}=160$:
$m_6=5.84$;

stage 7), lime of the 1$^{st}$ class of quality, $\Delta = 10$, $a_{st}=61$, $a_{fin}=200$:
$m_7=4.96$;

*stage* 8), lime of the 1$^{st}$ class of quality, $\Delta = 10$, $a_{st}=61$, $a_{fin}=160$, spruce of the 1$^{st}$ class of quality, $\Delta = 10$, $a_{st}=61$, $a_{fin}=160$:
$m_8=4.83$;

*stage* 9), lime of the 1$^{st}$ class of quality, $\Delta = 10$, $a_{st}=1$, $a_{fin}=160$, spruce of the 1$^{st}$ class of quality, $\Delta = 10$, $a_{st}=1$, $a_{fin}=160$:
$m_9=31.8$.

Finally, the vector $M_9=(m_1,\ldots m_9)=(15.5,\ 11.82,\ 11.23,\ 11.23,\ 11.23,\ 5.84,\ 4.96,\ 4.83,\ 31.8)$.

We will consider also shortened vector for the model with 8 stages of successions: $M_8=(15.5,\ 11.82,\ 11.23,\ 11.23,\ 11.23,\ 5.84,\ 4.96,\ 4.83)$.

*Remark*. We have no tables of growth for mixed lime and spruce forests; the simple way to overcome this problem is to "mix" these breeds in some given proportion. Using expert estimations (Korotkov, private communication) we took 40% of spruce and 60% of lime for computation of $m_8$ and $m_9$. More sophisticated approach could be based on analytical models of growth of two-breed forest. We have constructed this variant of DMTG; the computer experiments showed that, as a rule, it is possible to consider the growth of both breeds *independently* on each other without loss of accuracy. The model allows to take into account the impact of one breed to the growth of another as a result of the competition for light, but it does not increase essentially the accuracy of the model solution (but severely increase the model complexity).

Thus, if to put $V$=100(%), the mean time of renewing $m = \sum_{j=1}^{n} u_j m_j$ and the ergodic constant $K$=100/$m$ for different succession matrixes is: for succession matrix $\Gamma_1$ with 9 stages of succession $m(\Gamma_1)$=17.05; $K(\Gamma_1)$=5.87; for succession matrix $\Gamma_2$ with 8 stages of succession $m(\Gamma_2)$= 7.585, $K(\Gamma_2)$=13.18;

We can see that there exists essential difference in the values of $m$ and $K$ between the two versions: the version A) with 9 stages including the last ("climax") stage has severely smaller values of ergodic constant then the version B) with 8 stages.

Now we are able to apply the graphical "ergodic method" described in s. 4.3.

### 4.5.3. *The Stationary Distribution of Areas*

Let us consider on the plain $(T, S)$ the set of points $\mathbf{S}(\Gamma)$={$(T_j, S_j)$}, where $T_j = u_j m_j$ is the "proper time" of the $j$-th stage corresponding to the succession matrix $\Gamma$, and $S_j$ is the real area (in %) occupied by the $j$-th stage according Table 4.1.

For the succession matrix $\Gamma_1$ we have got the following values of proper times $\mathbf{T}(\Gamma_1)$={$T_j = u_j m_j, j$=1,…9}={0.461, 0.703, 0.483, 0.635, 0.584, 0.493, 0.77, 0.492, 12.4} and for the matrix $\Gamma_2$

$\mathbf{T}(\Gamma_2)$={$T_j = u_j m_j, j$=1,…8}={0.756, 1.154, 0.793, 1.04, 0.958, 0.809, 1.264, 0.808}

Let us take the real data for the Prioksko-Terrasnyi Biosphere Reserve in 1999 from the following Table (V.Korotkov, private communication):

**Table 4.2. Prioksko-Terrasnyi Biosphere Reserve in 1999**

| № of stage | Area, ha | Portion of area, % |
|---|---|---|
| 1 | 1.2 | 0.1 |
| 2 | 3.3 | 0.3 |
| 3 | 293.2 | 26.0 |
| 4 | 385.4 | 34.2 |
| 5 | 353.8 | 31.4 |
| 6 | 135.2 | 10.8 |
| 7 | 24.5 | 2.0 |

| 8 | 56.5 | 4.5 |
| Total | 1253.1 | 100.0 |

Let us define the set $\mathbf{S}(\Gamma_1;1999)=\{(T_j,S_j)\}$ and $\mathbf{S}(\Gamma_2;1999)=\{(T_j,S_j)\}$ for matrix $\Gamma_1$ and $\Gamma_2$ correspondingly and data for 1999 year,

$\mathbf{S}(\Gamma_1;1999)=\{\{0.461,\ 0.1\},\ \{0.703,\ 0.3\},\ \{0.483,\ 26.0\},\ \{0.635,\ 34.2\},\ \{0.584,\ 31.4\},$
$\{0.493,\ 10.8\},\ \{0.77,\ 2.0\},\ \{0.492,\ 4.5\},\ \{12.4,\ 0\}\};$

$\mathbf{S}(\Gamma_2;1999)=\{\{0.756,\ 0.1\},\ \{1.154,\ 0.3\},\ \{0.793,\ 26.0\},\ \{1.04,\ 34.2\},\ \{0.958,\ 31.4\},$
$\{0.809,\ 10.8\},\ \{1.264,\ 2.0\},\ \{0.808,\ 4.5\}\}.$

If a condition of the forest system would be close to the stationary state, then the corresponding set of points should lay close to the strait line with the slope $K(\Gamma_1)=5.87$ and $K(\Gamma_2)=13.18$ accordingly. In more detail, the theoretical proportions of stationary areas for the 9-stages succession scheme should be $\Theta(\Gamma_1)=\{5.87*\ T_j,\ j=1,\ldots9\}$ and for the 8-stages succession scheme $\Theta(\Gamma_2)=\{13.18*\ T_j,\ j=1,\ldots8\}$. Let us consider both cases.

The points $\{T_j,\ 5.87*T_j\ \},\ j=1,\ldots9$ are marked on the line $K(\Gamma_1)$, Fig.4.2 (the last point $\{12.6,\ 73.17\}$ is out of the plot).

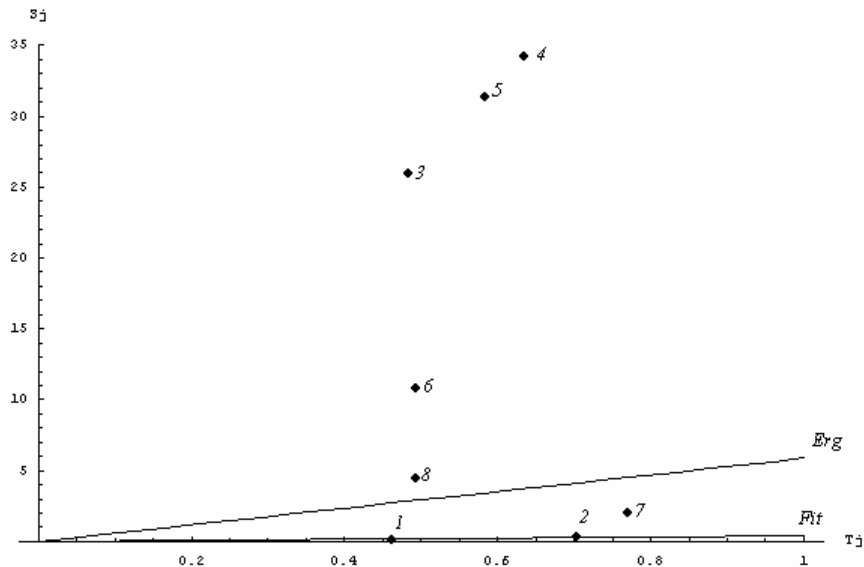

**Fig.4.2.** The model with 9 succession stages (matrix $\Gamma_1$). Real data for 1999 of the Prioksko-Terrasnyi Biosphere Reserve and theoretical stationary states; *Erg*-the line $S=5.87T$, which corresponds to theoretical values of areas of 9 stages in the stationary state; *Fit*-the line $S=0.396T$ (the point for the last 9[th] stage with $T_9=12.6$ is out of the plot)

It follows from the ergodic theorem, that the mean-square deviation of the set of points $\mathbf{S}=\{(T_j,S_j)\}$ from the "ergodic line" $S=KT$ tends to zero as the system's condition tends to the stationary state. Thus let us find the least-squares fit to the given set $\mathbf{S}(\Gamma_1;1999)$ by a line $S=kT$: the best-fit line is $S=k_1T$ with $k_1=0.396$. The reason of so significant difference between $k_1$ and $K$ is obvious: the set $\mathbf{S}(\Gamma_1;1999)$ contains the last point $\{12.6,\ 0\}$, which corresponds to the "absent" 9[th] stage. Theoretical portion of the area of the 9[th] stage should be the largest one and equal to 12.6*5.87=73.17, on the contrary with the zero data.

We can see (Fig. 4.2) that real proportions of areas are dramatically different from the theoretical proportions in the stationary state, which should lie on the line *Erg*: $S$=5.87$T$. We could conclude that the data on forest areas of the Prioksko-Terrasnyi Biosphere Reserve in 1999 do not correspond to succession schemes having 9 stages so that the 9-stages scheme is inappropriate. The same conclusion is valid for data on forest areas in 1955 (see below the values and fit of these data).

Next, let us consider the shortened version of the model with 8 stages and the succession matrix $\Gamma_2$. In this case the ergodic line is $S$=13.18$T$. Let us mark on Fig.4.3 the set of points $\{T_j, 13.18*T_j\}$, $j$=1,…8, which correspond to the theoretical stationary areas on the line $K(\Gamma_2)$, and the set of points $\mathbf{S}(\Gamma_2;1999)$, which correspond to real data from Table 4.2.

If the conditions of the forest system would be close to the stationary state, this set of points should lie close to the ergodic strait line with the slope $K(\Gamma_2)$=13.18. A least-squares fit to the given set $\mathbf{S}(\Gamma_2;1999)$ by a line $S$=$kT$ is $S$=$k_2T$ with $k_2$=13.66. We can see that the *ergodic* line *Erg* and *Fit*-line on Fig.4.3 are very close. It means that the real data on area proportions do not contradict to the succession scheme with 8 stages described by matrix $\Gamma_2$ (on the contrary with succession schemes with 9 stages).

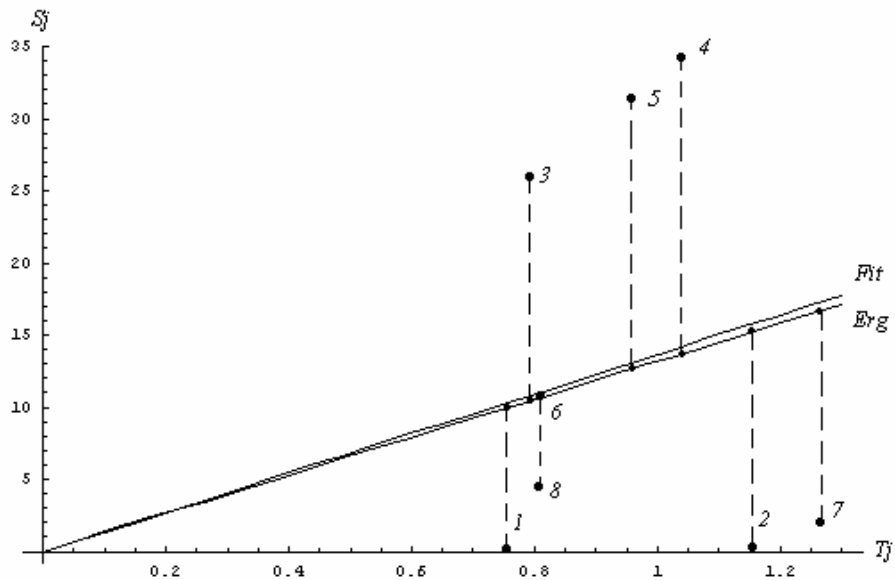

**Fig.4.3.** The model with 8 succession stages; theoretical stationary states and real data in 1999 of the Prioksko-Terrasnyi Biosphere Reserve. The dash lines show the points on the line *Erg*: $S$=13.18$T$ corresponding to the theoretical stationary values of areas for the 8-stages succession scheme with matrix $\Gamma_2$; the line *Fit*: $S$=13.66 $T$

Next, let us consider the data on forest areas of the Prioksko-Terrasnyi Biosphere Reserve in 1955 and apply the ergodic method to these data.

**Table 4.3. Prioksko-Terrasnyi Biosphere Reserve in 1955**

| № of stage | Area, ha | Portion of area, % |
|---|---|---|
| 1 | 374 | 26.3 |
| 2 | 508.3 | 35.8 |
| 3 | 52.9 | 3.7 |
| 4 | 144.3 | 10.2 |
| 5 | 250.9 | 17.7 |
| 6 | 61.1 | 4.3 |
| 7 | 6.9 | 0.5 |
| 8 | 21.1 | 1.5 |
| Total | 1419.5 | 100.0 |

Let us define the set $\mathbf{S}(\Gamma_2;1955)=\{(T_j,S_j)\}$ for matrix $\Gamma_2$ (which corresponds to 8-stage succession scheme) and data in 1955 year,

$\mathbf{S}(\Gamma_2;1955)$ =\{\{0.756, 26.3\}, \{1.154, 35.8\}, \{0.793, 3.7\}, \{1.04, 10.2\}, \{0.958, 17.7\}, \{0.809, 4.3\}, \{1.264, 0.5\}, \{0.808, 1.5\}\}.

A least-squares fit to the set $\mathbf{S}(\Gamma_2;1955)$ by a line $S=kT$ is $S=k_2T$ with $k_2=13.06$. We can see again that the *ergodic* line $S=13.18T$ and *Fit*-line $S=13.06T$ are very close, see Fig.4.4 where these lines almost coincide.

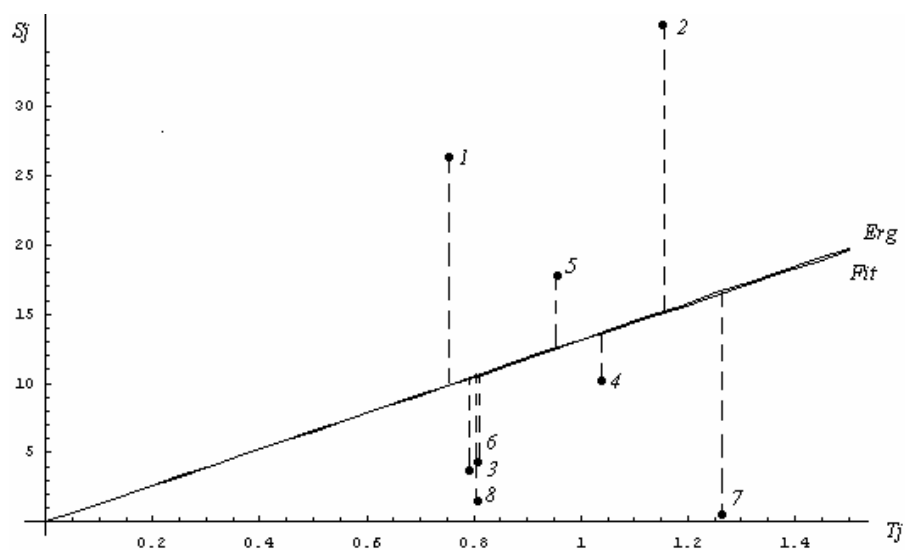

**Fig.4.4.** Theoretical stationary states and real data in 1955 for the Prioksko-Terrasnyi Biosphere Reserve. The dash lines show the points on the line *Erg*: $S=13.18T$ corresponding to the theoretical values of stationary areas of the 8-stages succession scheme with matrix $\Gamma_2$; *Fit*-the line $S=13.05T$.

Therefore, we may conclude that existing real data for Prioksko-Terrasnyi Biosphere Reserve correspond to 8-stages succession scheme with matrix $\Gamma_2$. Nevertheless, we observe comparatively large deviations of data points in both 1955 and 1999 from the theoretical stationary state. It is interesting that a strong reorganization of the area distribution was happening between 1955 and 1999, as we can see in Tables and Fig.-s 4.2-4.3. For example, the 1[st] and the 2[nd] succession stages dominated in 1955 and later transited to the 3[rd] and the 4[th] stages accordingly but practically were not renewed at 1999. It reflects the real history for the Prioksko-Terrasnyi Biosphere Reserve that was organized in 1945 on the territory with large fire-sites and cleared space after cutting down and battles of the War II.

If the intensities of succession transitions would be fixed (as in matrix $\Gamma_2$) then in course of time (approximately after 350-400 years) the portions of area occupied by different forest stages become very close to the theoretical stationary values, which we estimated as $\Theta(\Gamma_2)=\{13.18* T_j, j=1,\ldots8\}=\{9.97, 15.21, 10.45, 13.71, 12.63, 10.67, 16.66, 10.65\}$.

By the same way we can compare the real and theoretical stationary portions $S_j(a)$ of area occupied by the forests of given age at different stages. Corresponding "proper times for age area" are $T_j(a)= u_j(1-\exp(-N_j(a)))$, according formula (4.7). Recall that for *any* age *a* the set of points $\{T_j(a), S_j(a)\}$ belong to the *same ergodic line*. The same comparison could be done for real and theoretical stationary values of wood stock for every succession stage.

#### 4.5.4. *Management in Dankov Forestry*

Let us consider a simple case of forest management on the example of Dankov forestry, Moscow region, Russia. Approximately 0.5%-0.7% of forest areas is cut off every year in this forestry. We will suppose in what follows that the area of every stage (except the 1[st] and the 2[nd] ones) is cut off with the intensity 0.6% per year.

The stages and their durations under natural course of succession are the same as given in Fig.4.1 and Table 4.1. Detailed investigation of different successive schemes performed above is applicable to the Dankov forestry also and showed that the best variant is the successive scheme with 8 successive stages.

The real data for the forestry are given in the following Table (V.Korotkov, private communication).

#### Table 4.4. Dankov forestry, 1990

| № of stage | Area, ha | Portion of area, % |
|------------|----------|--------------------|
| 1          | 4.2      | 0.2                |
| 2          | 24.6     | 1.1                |
| 3          | 1145.8   | 50.0               |
| 4          | 550.9    | 24.1               |
| 5          | 216.6    | 9.5                |
| 6          | 166.9    | 7.3                |
| 7          | 54.2     | 2.4                |
| 8          | 127.4    | 5.6                |
| Total      | 2290.6   | 100.0              |

As the stages and their durations under natural course of succession are the same as for the Prioksko-Terrasnyi Biosphere Reserve, we could define the succession matrix by the same way as the matrix $\Gamma_2$ for the Reserve with the following differences. At first, let us suppose again (as it was done for the matrix $\Gamma_2$) that the free area after the 2$^{nd}$ stage is distributed between the 3-5 stages proportionally to the areas occupied by these stages according to the data in 1990, i.e. 50 : 24.1 : 9.5 (see Table 4.4). Secondly, we take into account a management in the simplest form: we suppose that 0.6% of area of the stages 3-8 is cut off every year and the free areas add to the first stage. Then we have the following succession matrix $\Gamma_3 = P3^T$:

$$\Gamma_3 = \begin{pmatrix}
0.9 & 0. & 0.006 & 0.006 & 0.006 & 0.016 & 0.013 & 0.016 \\
0.1 & 0.95 & 0. & 0. & 0. & 0. & 0. & 0. \\
0. & 0.03 & 0.98 & 0. & 0. & 0. & 0. & 0. \\
0. & 0.014 & 0. & 0.98 & 0. & 0. & 0. & 0. \\
0. & 0.006 & 0. & 0. & 0.98 & 0. & 0. & 0. \\
0. & 0. & 0.017 & 0. & 0. & 0.98 & 0. & 0. \\
0. & 0. & 0. & 0.017 & 0. & 0. & 0.99 & 0. \\
0. & 0. & 0. & 0. & 0.017 & 0. & 0. & 0.98
\end{pmatrix}$$

This matrix has the eigenvector (corresponding to the eigenvalue 1): $\boldsymbol{U_3} = (u_1, \dots u_8) = $ (0.081, 0.163, 0.215, 0.103, 0.041, 0.224, 0.131, 0.042). The values $\{m_j, j=1,\dots 8\}$ are the same as for Prioksko-Terrasnyi Biosphere Reserve and so the ergodic constant $K(\Gamma_3) = 1.41$.

We have got the following values of proper times for the succession matrix $\Gamma_3$: $T(\Gamma_3) = \{T_j = u_j m_j, j=1,\dots 8\} = \{1.26, 1.92, 2.05, 0.99, 0.39, 1.31, 0.65, 0.21\}$.

Let us define the set $\mathbf{S}(\Gamma_3; 1990) = \{(T_j, S_j)\}$ for the matrix $\Gamma_3$ and the data from Table 4.4:

$\mathbf{S}(\Gamma_3; 1990) = \{\{1.26, 0.2\}, \{1.92, 1.1\}, \{2.05, 50.\}, \{0.99, 24.1\}, \{0.39, 9.5\}, \{1.31, 7.3\}, \{0.65, 2.4\}, \{0.21, 5.6\}\}$.

A least-squares fit to the set $\mathbf{S}(\Gamma_3; 1990)$ by a line $S = kT$ is $S = k_4 T$ with $k_4 = 11.31$. We can see again that the *ergodic* line $S = 11.41T$ and *Fit*-line $S = 11.31T$ are very close, see Fig.4.5 where these lines almost coincide. It means that the model with the matrix $\Gamma_3$ accounted the natural course of succession together with the management does not contradict the real data and could be applied for the prognoses of the stationary condition of the forestry under different variant of management.

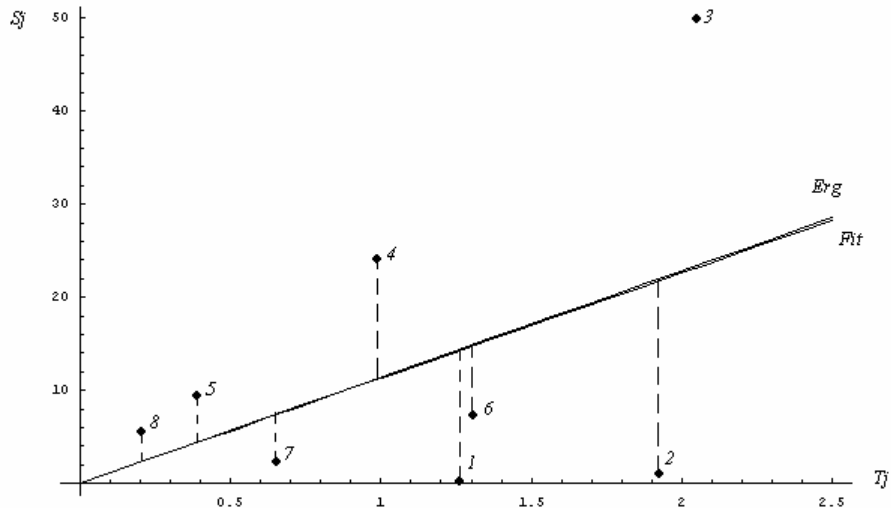

**Fig.4.5.** Theoretical stationary state and the real data for the Dankov forestry, 1990. The dash lines show the points on the line *Erg*: $S=11.41T$ corresponding to the theoretical values of areas of 8-stages succession scheme with matrix $\Gamma_3$; *Fit* - the line $S=11.31T$

### 4.5.5. *The Stationary Distributions of the Store of Wood and other Generalized Variables*

Studying of the stationary distribution of areas being closely connected with the initial formulation of the ergodic hypothesis is mainly of theoretical interest. For large territories, the area distribution could be roughly estimated with the help of aero photos. Nevertheless, although the notion of area $S_i(t,a,X)$ is a very useful mathematical abstraction it could not be correlated with a measurable value in many real situations. On the contrary, the store of wood (and its distribution) is the most interesting and easy measured value. To calculate the stationary distribution of the store of wood we should use formula (4.16),

$$W_j /[\, u_j \int_o^\infty \; W_j(a)(1- \exp(-N_j(a)))\; da\,] = K.$$

According to the theory developed above, the value of the ergodic constant $K$ is exactly the same as it was calculated for the stationary distribution of areas in the previous section.

Thus we get the following algorithm for computation of the stationary values of wood stock in all succession stages and comparing the real data with the theoretical stationary ones.

First of all, the succession matrix $\{\Gamma_{ij}\}$ should be defined and its left eigenvector ($u_1$, ... $u_n$) corresponding to the eigenvalue 1 should be computed. According to the previous investigation, we choose the matrix $\Gamma_2$. The following steps are:

  (i).       to compute the values $w_i = \int\limits_{o}^{\infty} W_j(a)(1-\exp(-N_j(a)))\mathrm{d}a$;

  (ii).      according to formula (4.12), to define the "proper times for wood stock", $WT_i = u_i w_i$ for each stage $i = 1, ... k$ ;

  (iii).     on the "ergodic line" $y = Kx$ (with $x = WT$, the proper time, and $y = W$, the stock) to postpone the points with $x$-coordinates $WT_i$, $i = 1, ... k$, which show the theoretical stationary values of the wood stock of corresponding stages;

  (iv).      to put on the plane $(x,y)$ (or $(WT,W)$) the set of points $\{W_i(t), WT_i, i = 1, ... k\}$, where $W_i(t)$ is the wood stock of the $i$-th stage at time $t$ and to estimate the deviation of this set of points from the ergodic line $y = Kx$.

The problem of computation of the values $w_i$ could be rather difficult. To make possible a correct comparisons with real data the values $W_j(a)$ should be computed with the help of exactly verified model of mixed forest growth. Such models are rather bulky (Franc et al. 1995, 2000; Porte, Baterlink 2001) and the problem of theirs "tuning" is difficult. Another, simplified method of computation of the values $w_i$ is based on direct using of Forest Growth Tables and should take into account the composition and other specific characteristics of the studying forest.

Detailed investigation of the stationary distribution of wood stock and other generalized variables for specific forests is beyond of this paper.

## CONCLUSION

The succession models compose the top level in the hierarchical system of models intended as for understanding of natural dynamics of untouched forests so for estimation and prediction of the forest ecosystem conditions under any "stationary" management. We have shown that the developed *ergodic method* of estimation of real forest conditions based on the structural models of succession and the ergodic theorem, suggests comparatively simple algorithm for computation of the stationary state of the forest system under given scenario of the succession transition and management policy. The results should be compared with real data for estimation how close the current forest conditions are to the stationary ones. The method could help to choose the desirable management policy.

Acknowledgment. The research was partly supported by INTAS grant 97-30255. The author is very appreciating to Dr. V. Korotkov for some data and very useful comments and discussions.

# MATHEMATICAL APPENDIX

## 1. Markov's Renewal Equations

The classical renewal theory studies the asymptotic behaviors of the solutions to the renewal equation

$$Z(t)=f(t)+\int_0^t M(\mathrm{d}u)Z(t\text{-}u)$$

(A.1)

or to the matrix renewal equation

$$Z_i(t)=f_i(t)+\int_0^t \sum_{j=1}^n M_{ij}(\mathrm{d}u)Z(t\text{-}u).$$

(A.2)

Markov's renewal theory studies the systems of functional equations

$$Z_i(t,x)=f_i(t,x)+\int_\Omega\int_0^t \sum_{j=1}^n M_{ij}(x,\mathrm{d}u,\mathrm{d}y)Z_j(t\text{-}u,y)$$

(A.3)

where the vector-function $f=(f_1,...f_n)$ and the matrix kernel $M=(M_{ij})$ are given.

The basic results about asymptotic behaviors of the solution to the functional Markov equation for the case $n = 1$ are given in (Shurenkov, 1989). This theory is rather difficult and the checking of the theorem's conditions may be a non-trivial problem. Here we show that the theory of systems of the Markov renewal equations (3) for any $n$, corresponding to special kernels $M$ can be completely reduced to the theory of more simple matrix renewal equations (A.2).

Let us formulate some known results for equations (A.2). Consider the matrix $M(\infty)=\left| M_{ij}(\infty)\right|$ and suppose that it is *indecomposable*. Let $\lambda$ be the Perron's root of the matrix $M(\infty)$. It is known (see, e.g., Kovalenko et al., 1986, ch.12.6) that if the root $\lambda>1$, than there exists the unique positive number $\mu$ such that the Perron's root of the matrix

$$M^{\#}_{ij}=\int_0^\infty \exp(\text{-}t\mu)M_{ij}(\mathrm{d}t) \text{ is equal to 1.}$$

If $\lambda <1$ then the corresponding (negative) $\mu$ exists if and only if $\int_0^\infty \exp(\text{-}t\varepsilon)M_{ij}(\mathrm{d}t)<\infty$ for all $i,j$ and some $\varepsilon >0$. Multiplying both sides of this equation to $\exp(\text{-}t\mu)$, we get the matrix renewal equation

$$Z^{\#}_i(t)=f^{\#}_i(t)+\int_0^t \sum_{j=1}^n M^{\#}_{ij}(\mathrm{d}u)Z^{\#}_j(t\text{-}u)$$

where $Z^{\#}_i(t) = Z_i(t) \exp(-t\mu)$, $f^{\#}_i(t) = f_i(t) \exp(-t\mu)$, and the Perron's root of the matrix $M^{\#}_{ij}$ is equal to 1.

*Definition*. A function $f(t)$ is *directly integrable by Riman on* $[0,\infty)$, if

$$\sum_{n\geq 0} \sup_{n\leq t \leq n+1} |f(t)| < \infty,$$

$$\delta \sum_{n\geq 0} \{\sup_{n\delta \leq t \leq n\delta+1} f(t) - \inf_{n\delta \leq t \leq n\delta+1} f(t)\} \to 0 \ under \ \delta \to 0.$$

**Theorem 1** (*The Smith's renewal theorem, Kovalenko et al., 1986, theorem* 12.38). *Let the matrix* $M(\infty) = |M_{ij}(\infty)|$ *is indecomposable and its Perron's root is equal to* 1 *and the matrix* $M(t) = |M_{ij}(t)|$ *is non-latticed. Let* $(v_1,...v_n)$, $(u_1,...u_n)$ *be the left and the right eigenvectors of the matrix* $M(\infty)$, *corresponding to the eigenvalue* 1. *If*

$$0 < m = \sum_{i,j=1}^{n} v_i u_j \int_0^{\infty} t\, M_{ij}(\mathrm{d}t) < \infty$$

*and the functions* $f_i(t)$, $i=1,...n$ *are directly integrable by Riman on* $[0,\infty)$, *then*

$$\lim_{t\to\infty} Z_i(t) = u_i V / m$$

$$where \ V = \sum_{j=1}^{n} v_j \int_0^{\infty} f_j(t)\mathrm{d}t.$$

*Remark*. Under some additional conditions on the matrix $M(\infty)$ it is enough to assume, that the functions $f_i(t)$ are bounded, integrable on $[0,\infty)$ and tend to 0 at $t\to\infty$, see Kovalenko et al., 1986, theorem 12.36.

Now let us consider the functional Markov equation (3). Let $(\Omega, A)$ be a measurable space where $\Omega$ is a domain in $R^n$. Consider measurable space $(\Omega \times [0,\infty), A \times B)$ where $B$ is the Borel' $\sigma$- algebra on $[0,\infty)$. Let the non-negative matrix kernel $\boldsymbol{M} = |M_{i,j}(x,\mathrm{d}y,\mathrm{d}t)|$ is given on this measurable space. By definition, the matrix kernel $\boldsymbol{M}$ is an $A$-measurable by $x$ measure $M_{i,j}(x,\mathrm{d}y,\mathrm{d}t)$ on $(A \times B)$ for each $i,j = 1,...n$ such that

$$\sup_x M_{i,j}(x, \Omega, [0,\infty)) < \infty.$$

The resolvent of the kernel $\boldsymbol{M}$ is the kernel $\boldsymbol{R} = |R_{ij}|$ where

$$R_{ij}(x,\mathrm{d}t,\mathrm{d}y) = \sum_{k=0}^{\infty} M_{ij}^{(k)}(x,\mathrm{d}t,dy).$$

Here $M_{ij}^{(k)}$ is the k-times convolution of the kernel $M_{ij}$, i.e.

$$M_{ij}^{(0)}(x,[0,t),A) = \begin{cases} 1, x \in A \\ 0, x \notin A \end{cases},$$

$$M_{ij}^{(k+1)}(x,[0,t),A) = \iint\limits_{\Omega\ 0} \sum_{s=1}^{n} M_{is}(x,\mathrm{d}u,\mathrm{d}y)\ M_{sj}^{(k)}(y,[0,t-u),A).$$

Let us denote $\boldsymbol{M}*\boldsymbol{f}$ the convolution of the kernel $\boldsymbol{M}$ and the vector - function $\boldsymbol{f}$, where

$$(\boldsymbol{M}*\boldsymbol{f})_i\ (t,x) = \iint\limits_{\Omega\ 0} \sum_{j=1}^{n} M_{ij}\ (x,\mathrm{d}u,\mathrm{d}y)\ f_j(t-u,y).$$

The unique solution of equation (3) is the convolution of the resolvent $\boldsymbol{R}$ and the vector - function $\boldsymbol{f}$:

$$Z(t,x) = (\boldsymbol{R}*\boldsymbol{f})(t,x). \qquad\qquad (A.4)$$

In what follows we suppose that $M_{ij}\ (x,\mathrm{d}u,\mathrm{d}y)$ read

$$M_{ij}\ (x,\mathrm{d}u,\mathrm{d}y) = \gamma_{ij}(x)\ U_j\ (u,y)\ \mathrm{d}u\mathrm{d}y \qquad\qquad (A.5)$$

for all $i,j$ and the functions $\gamma_{ij}$, $U_j$ are measurable and nonnegative,

$$\iint\limits_{\Omega\ 0}^{\ \infty} U_j(u,y)\ \mathrm{d}u\mathrm{d}y < \infty,\ \sup_x \gamma_{ij}(x) < \infty,\ \int\limits_{\Omega} \gamma_{ij}(x)\mathrm{d}x < \infty. \qquad\qquad (A.6)$$

Let us consider the system

$$Z_i(t,x) = f_i(t,x) + \sum_{j=1}^{n} \gamma_{ij}(x) \iint\limits_{\Omega\ 0} U_j(u,y)\ Z_j(y,t-u)\ \mathrm{d}u\mathrm{d}y \qquad\qquad (A.7)$$

where $f_1,\ldots f_n$ are given nonnegative functions.

The "characteristic number" $\lambda$ of kernel $\boldsymbol{M}$ (5) is defined by the following condition: the matrix

$$\Phi_{ij} = \iint\limits_{\Omega\ 0}^{\ \infty} \exp(-\lambda u)\ U_i(u,y)\ \gamma_{ij}(y)\mathrm{d}u\mathrm{d}y$$

is finite, indecomposable and its Perron' root is equal to 1.

We will suppose that the characteristic number $\lambda$ of the kernel $\boldsymbol{M}$ exists and $\lambda \geq 0$. (A.8)

Let us denote

$$N_i(t,x) = \exp(-\lambda t)U_i(t,x),$$
$$K_{ij}\ (x,t,y) = \gamma_{ij}(x)N_j(t,y),$$

$g_i(t,x) = \exp(-\lambda t) f_i(t,x),$

$W_i(t,x) = \exp(-\lambda t) Z_i(t,x),$

$F_{ij}(t) = \iint\limits_{\Omega} \int\limits_{0}^{t} N_i(u,y) \, \gamma_{ij}(y) \mathrm{d}u\mathrm{d}y,$

$\Phi_{ij} = F_{ij}(\infty).$

System (A.7) is obviously equivalent to the system

$$W_i(t,x) = g_i(t,x) + \sum_{j=1}^{n} \gamma_{ij}(x) \iint\limits_{\Omega}\int\limits_{0}^{t} N_j(u,y) \, W_j(y,t\text{-}u) \, \mathrm{d}u\mathrm{d}y \qquad \text{(A.9)}$$

and the characteristic number of the kernel $K_{ij}(x,t,y) = \gamma_{ij}(x)N_j(t,y)$ is equal to 0.

Let $(v_1,\ldots v_n)$, $(u_1,\ldots u_n)$ be unique (up to a constant factor) non-negative the left and the right respectively eigenvectors of the matrix $\Phi_{ij}$, corresponding to the eigenvalue 1; let us assume, that

$$\sum_{i=1}^{n} u_i v_i = 1.$$

Denote

$$m = \sum_{i,j=1}^{n} u_j v_i \int\limits_{0}^{\infty} t \, \exp(-\lambda t) \, dF_{ij}(t), \qquad \text{(A.10)}$$

$$\Gamma_i(y) = \int\limits_{0}^{\infty} \exp(-\lambda u) N_i(u,y) \mathrm{d}u,$$

$$V = \sum_{j=1}^{n} v_j \int\limits_{\Omega} \Gamma_j(y) \int\limits_{0}^{\infty} \exp(-\lambda t) f_j(t,y) \, \mathrm{d}t \, \mathrm{d}y \qquad \text{(A.11)}$$

Suppose that

$$m < \infty, \; V < \infty. \qquad \text{(A.12)}$$

At last, let us suppose that functions

$$h_j(t) = \exp(-\lambda t) \iint\limits_{\Omega}\int\limits_{0}^{t} U_j(u,y) \, f_j(y,t\text{-}u) \, \mathrm{d}u\mathrm{d}y$$

are *directly integrable by Riman on* $[0,\infty)$ for all $j$. \qquad \text{(A.13)}

The following theorem describes the asymptotic behavior of the solution of equation (A.7).

**Theorem 2**. *Under conditions* (A.8), (A.12), (A.13)

$$Z_i(t,x)\exp(-\lambda t) \to V/m \sum_{j=1}^{n} \gamma_{ij}(x)u_j$$

*at* $t \to \infty$ *for all* $i = 1,...n$.

Proof. The proof of this theorem is based on the following lemma.

Let us denote $\chi(x,y) = \begin{cases} 1, x = y \\ 0, x \neq y \end{cases}$

**Lemma**. *The resolvent* **R** *of the kernel* **K** *can be written in the form*

$$R_{ij}(x,t,y) = \chi(x,y) + \sum_{k=1}^{n} \gamma_{ik}(x) P_{kj}(t,y), \tag{A.14}$$

where the functions $P_{ij}(t,y)$ satisfy to the system

$$P_{ij}(t,y) = \chi(i,j) N_j(t,y) + \sum_{k=1}^{n} \int_{0}^{t} F_{ik}(u) P_{kj}(t-u,y)\mathrm{d}u \tag{A.15}$$

Proof of the lemma

Let us define $p_{ij}^{(m)}(t,y)$ by induction:

$$p_{ij}^{(1)}(t,y) = \chi(i,j)N_j(t,y),$$

$$p_{ij}^{(m+1)}(t,y) = \sum_{k=1}^{n} \int_{0}^{t} F_{ik}(u)p_{kj}^{(m)}(t-u,y)\mathrm{d}u.$$

Denote

$$P_{ij}(t,y) = \sum_{m=1}^{\infty} p_{ij}^{(m)}(t,y)$$

Obviously $P_{ij}(t,y)$ satisfy to equation (15). Now let us prove by induction that for all $m$

$$K_{ij}^{(m)}(x,t,y) = \sum_{k=1}^{n} \gamma_{ik}(x) p_{kj}^{(m)}(t,y).$$

Indeed, $K_{ij}^{(m+1)}(x,t,y) = \iint\limits_{\Omega} \int_{0}^{t} \sum_{k=1}^{n} \gamma_{ik}(x) N_k(t-u,z) K_{kj}^{(m)}(z,u,y) \,\mathrm{d}u\mathrm{d}z =$

$$= \iint\limits_{\Omega} \int_{0}^{t} \sum_{k=1}^{n} \gamma_{ik}(x) N_k(u,z) \sum_{s=1}^{n} \gamma_{ks}(z) p_{sj}^{(m)}(t-u,y) \,\mathrm{d}u\mathrm{d}z =$$

$$= \sum_{k,s=1}^{n} \gamma_{ik}(x) \int_{0}^{t} F_{ks}(u) p_{sj}^{(m)}(t-u,y) \,\mathrm{d}u = \sum_{k=1}^{n} \gamma_{ik}(x) p_{kj}^{(m+1)}(t,y).$$

Hence, $R_{ij}(x,t,y) = \sum\limits_{k=0}^{\infty} K_{ij}^{(k)}(x,t,y) = K_{ij}^{(0)}(x,t,y) + \sum\limits_{m=1}^{\infty}\sum\limits_{k=1}^{n} \gamma_{ik}(x)\, p_{kj}^{(m)}(t,y) =$

$= \chi(x,y) + \sum\limits_{k=1}^{n} \gamma_{ik}(x)\, P_{kj}(t,y)$, as desired.

Let us return to the proof of theorem 2 and denote

$$G_{ij}(t,y) = \int\limits_{0}^{t} P_{ij}(u,y) g_j(t-u,y)\mathrm{d}u$$

Then

$$G_{ij}(t,y) = \chi(i,j)\int\limits_{0}^{t} N_j(u,y) g_j(t-u,y)\mathrm{d}u + \sum\limits_{k=1}^{n}\int\limits_{0}^{t} F_{ik}(v) G_{kj}(t-v,y)\mathrm{d}v$$

(A.16)

Indeed,

$$P_{ij}(t,y) = \sum\limits_{k=1}^{\infty} p_{ij}^{(k)}(t,y) = \chi(i,j)\, N_j(u,y) + \sum\limits_{k=1}^{n}\int\limits_{0}^{t} F_{ik}(v) P_{kj}(t-v,y)\mathrm{d}v;$$

$$\sum\limits_{k=1}^{n}\int\limits_{0}^{t}\int\limits_{0}^{t} F_{ik}(v) P_{kj}(u-v,y)\, g_j(t-v,y)\mathrm{d}u\ \mathrm{d}v =$$

$$= \sum\limits_{k=1}^{n}\int\limits_{0}^{t}(\int\limits_{v}^{t} F_{ik}(v) P_{kj}(u-v,y)\, g_j(t-v,y)\mathrm{d}u)\mathrm{d}v = \sum\limits_{k=1}^{n}\int\limits_{0}^{t} F_{ik}(v) G_{kj}(t-v,y)\mathrm{d}v,$$

and (A.16) is proved.

Let us denote

$$Q_{ij}(t) = \int\limits_{\Omega} G_{ij}(t,y)\mathrm{d}y,$$

then

$$Q_{ij}(t) = \chi(i,j)\iint\limits_{\Omega\ 0} N_j(u,y) g_j(t-u,y)\mathrm{d}u\mathrm{d}y + \sum\limits_{k=1}^{n}\int\limits_{0}^{t} F_{ik}(v) Q_{kj}(t-v)\mathrm{d}v$$

(A.17)

Thus, the functions $Q_{ij}(t)$ for every fixed $j$ satisfied to system (A.2) with

$$f_i(t,x) = \chi(i,j)\iint\limits_{\Omega\ 0} N_j(u,y) g_j(t-u,y)\mathrm{d}u\mathrm{d}y.$$

According to Theorem 1 there exist $\lim\limits_{t\to\infty} Q_{ij}(t) = Q_{ij}(\infty)$,

where $Q_{ij}(\infty) = u_i/m \sum_{k=1}^{n} v_k \chi(k,j) \int_0^{\infty} \int_{\Omega} \int_0^t N_j(u,y) g_j(t-u,y) \mathrm{d}u \mathrm{d}y \mathrm{d}t =$

$= u_i/m \, v_j \int_{\Omega} (\int_0^{\cdot} N_j(u,y) \, \mathrm{d}u) \, (\int_0^{\cdot} g_j(t,y) \mathrm{d}t) \, \mathrm{d}y = u_i/m \, v_j \int_{\Omega} \Gamma_j(y) \int_0^{\cdot} g_j(t,y) \mathrm{d}t \, \mathrm{d}y.$

According to lemma and formula (A.4)

$W_i(t,x) = \sum_{j=1}^{n} \int_{\Omega} \int_0^t R_{ij}(x,t,y) \, g_j(t-u,y) \mathrm{d}u \mathrm{d}y =$

$= \sum_{j=1}^{n} \sum_{s=1}^{n} \gamma_{is}(x) \int_{\Omega} \int_0^t P_{ij}(x,t,y) \, g_j(t-u,y) \mathrm{d}u \mathrm{d}y = \sum_{j=1}^{n} \sum_{s=1}^{n} \gamma_{is}(x) \, Q_{sj}(t),$

and hence

$\lim_{t \to \infty} W_i(t,x) = \sum_{j=1}^{n} \sum_{s=1}^{n} (\gamma_{is}(x)u_s / m) \, v_j \int_{\Omega} \Gamma_j(y) \int_0^{\cdot} g_j(t,y) \mathrm{d}t \, \mathrm{d}y = V/m \sum_{s=1}^{n} \gamma_{is}(x)u_s$

The theorem is proved.

## 2. Structural Models of Communities and the Ergodic Theorems

The structural model of community is set by the system

$$
\left.
\begin{aligned}
&dX_i / da = F_i(a, X) \\
&\partial l_i / \partial t + \partial l_i / \partial a + div(l_i F_i) = -\mu_i l_i \\
&l_i(t,0, X) \equiv B_i(t, X) = \sum_{k=1}^{n} \gamma_{ik}(X) \int_{\Omega} \int_0^{\infty} \beta_k(a,Y) l_k(t,a,Y) dY da \\
&l_i(0, a, X) = l_i^0(a, X)
\end{aligned}
\right\}
\qquad \text{(A.18)}
$$

where $i = 1,...n$, $X = (x_1,....x_m)$, $\Omega \subseteq R_n^+$ is the range of values of the vector $X$.

Model (A.18) is called non-linear (or non-autonomic) if the rates of growth, $F_i$, death, $\mu_i$, or birth, $\beta_i$ depend on "regulators" of the form

$$
Q_i(t) = \int_{\Omega} \int_0^{\infty} q(a, X) l_i(t, a, X) da dX
$$

or

$$G_i(a,t) = \int\limits_{\Omega} q(a,X)l_i(t,a,X)dX .$$

In the opposite case the model is called autonomic. General properties of non-linear structured models were investigated in the works (Gurtin, MacCamy, 1974, Webb, 1994, Tucker, Zimmerman, 1988, etc.). Below we consider only autonomic structure models (Karev, 1993, Cushing, 2001).

Assume that the functions $F_i$ are continuous and bounded together with its first partial derivations, the functions $\mu_i, \beta_k, \gamma_{ik}, l_i^0$ are nonnegative, continuous and bounded and

$$\int\limits_{\Omega}\int\limits_{0}^{\infty} l_i^0(a,X)\mathrm{d}a\mathrm{d}X < \infty, \int\limits_{\Omega} \gamma_{ik}(X)\mathrm{d}X < \infty. \tag{A.19}$$

The succession matrix $\Gamma_{ij} = \int\limits_{\Omega} \gamma_{ij}(X)\mathrm{d}X$ is stochastic by supposition and hence its right eigenvector $(v_1,....v_n)$ corresponding to the eigenvalue 1 is $(1,...1)$.

Let $X_i(a,Y)$ be the solution of the Cauchy problem

$\mathrm{d}X_i/\mathrm{d}a = F_i(a,X), X(0)=Y.$

Denote

$$n_i(a,Y) = \exp\left(-\int\limits_{0}^{a} \mu_i(u,X(u,Y))du\right)$$

$$J_i(a,Y) = \exp\left(-\int\limits_{0}^{a}\sum_{j=1}^{n} \partial F_{ij}/\partial x_j(u,X(u,Y))du\right)$$

Then it could be shown (similar to Tucker, Zimmerman, 1988) that

$$l_i(t,a,X(a,Y)) = l_i^0(a-t,X(a-t,Y)) \, n_i(a,Y)J_i(a,Y)/(n_i(a-t,Y)J_i(a-t,Y)) \text{ if } a \geq t, \tag{A.20}$$

$$l_i(t,a,X(a,Y)) = B_i(t-a,Y) \, n_i(a,Y)J_i(a,Y) \text{ if } a < t. \tag{A.21}$$

An important formula for regulating functionals follows from these relations:

$$Q_i(t) = \int\limits_{\Omega} \int\limits_{0}^{\infty} q(a,X)l_i(t,a,X)dadX =$$

$$g_i(t) + \int\limits_{\Omega} \int\limits_{0}^{t} q(a,X(a,Y))n_i(a,Y)B_i(t-a,Y)dadY \tag{A.22}$$

where

$$g_i(t) = \iint\limits_{\Omega}\int\limits_{t}^{\infty} q(a, X(a,Y))\, l_i^0\, (a\text{-}t,Y))\, \exp(-\int\limits_{a-t}^{a} \mu_i(u, X(u,Y))du\,)\mathrm{d}a\mathrm{d}Y \tag{A.23}$$

In particular, the functions

$$\zeta_i(t) = \iint\limits_{\Omega}\int\limits_{0}^{\infty} \beta_i(a,Y)l_i(t,a,Y)dYda$$

satisfy to the equations

$$\zeta_i(t) = b_i(t) + \int\limits_{\Omega}\int\limits_{0}^{t} \beta_i(a,X(a,Y))n_i(a,Y)B_i(t-a,Y)dadY \tag{A.24}$$

where

$$b_i(t) = \iint\limits_{\Omega}\int\limits_{t}^{\infty} \beta_i(a, X(a,Y))\, l_i^0\, (a\text{-}t,Y))\, \exp(-\int\limits_{a-t}^{a} \mu_i(u, X(u,Y))du\,)\mathrm{d}a\mathrm{d}Y .$$

So, the functions

$$B_i(t,X) = \sum_{k=1}^{n} \gamma_{ik}(X)\varsigma_k(t)$$

satisfy to the system of Markov renewal equations

$$B_i(t,x) = f_i(t,x) + \sum_{k=1}^{n} \gamma_{ij}(x)\int\limits_{\Omega}\int\limits_{0}^{t} \beta_k(a,X(a,Y))n_k(a,Y)B_k(t-a,Y)dadY \tag{A.25}$$

where

$$f_i(t,x) = \sum_{k=1}^{n} \gamma_{ik}(X)b_k(t)$$

Now we can apply Theorem 2 to system (25) and get the asymptotics of $B_i(t,x)$ at $t\rightarrow\infty$:

$$B_i(t,x)\exp(-\lambda t)\rightarrow V/m \sum_{j=1}^{n} \gamma_{ij}(x)u_j.$$

Using formula (A.21) we get an important conclusion:

$$l_i(t,a,X(a,Y))\exp(-\lambda t) \rightarrow V/m\,\exp(-\lambda a)\,n_i(a,Y)J_i(a,Y)\sum_{j=1}^{n}\gamma_{ij}(x)u_j$$

(A.26)

Let us consider in more detail a *balanced* (or *equilibrium*) model with $\beta_i=\mu_i$. Then

$$\mathrm{d}n_i(a,Y)/\mathrm{d}a = -\,n_i(a,Y)\mu_i(a,X(a,Y)) = -\,n_i(a,Y)\beta_i(a,X(a,Y))\,,\text{ so}$$

$$\int_0^t \beta_i(a,X(a,Y))n_i(a,Y)B_i(t-a,Y)da = -\int_0^t (dn_i(a,Y)/da)B_i(t-a,Y)da\,.$$

System (A.25) read now

$$B_i(t,x)=f_i(t,x)-\sum_{k=1}^{n}\gamma_{ij}(x)\int_\Omega\int_0^t (dn_i(a,Y)/da)B_i(t-a,Y)dadY$$

(A.27)

The matrix

$$\Phi_{ij}=-\int_\Omega\int_0^\infty (dn_i(u,Y)/du)\,\gamma_{ij}(Y)\mathrm{d}u dY=\int_\Omega \gamma_{ij}(Y)\mathrm{d}Y=\Gamma_{ij},$$

hence, by suppositions, it is finite, indecomposable and its Perron' root is equal to 1.

Now we can apply Theorem 2 and conclude that

$$B_i(t,Y)\rightarrow V/m\sum_{j=1}^{n}\gamma_{ij}(Y)\,u_j$$

(A.28)

at $t\rightarrow\infty$ for all $i=1,\dots n$.

Here

$$m=\sum_{i,j=1}^{n}u_j\int_0^\infty t\,dF_{ij}(t)=-\sum_{i,j=1}^{n}u_j\int_\Omega\int_0^\infty t\,(dn_i(t,Y)/dt)\,\gamma_{ij}(Y)\mathrm{d}Y=$$

$$\sum_{i,j=1}^{n}u_j\int_\Omega\int_0^\infty n_i(t,Y)\,\mathrm{d}t\,\gamma_{ij}(Y)\mathrm{d}Y=$$

$$=\sum_{i,j}u_j\int_\Omega m_i(Y)\gamma_{ij}(Y)\,dY$$

(A.29)

where

$$m_j(X) = \int\limits_0^\infty n_j(a,X)\,da$$

According to formula (A.11) the constant

$$V = \sum_{j=1}^n u_j \int\limits_\Omega \int\limits_0^\infty f_j(t,y)\,dt\,dy, \text{ as}$$

$$\Gamma_k(y) = \int\limits_0^\infty \beta_k(u,X(u,Y))n_k(u,Y)\,du = -\int\limits_0^\infty (dn_i(u,Y)/du)\,du = 1.$$

After some technical algebra, one can show that finally for model (A.18)

$$V = \sum_{j=1}^n \iint\limits_\Omega \int\limits_0^\infty l_j^0(a,X)\,da\,dX.$$

Now it follows from (A.28) and (A.21), that

$$l_i(t,a,X(a,Y)) \to V/m\, n_i(a,Y)J_i(a,Y) \sum_{j=1}^n \gamma_{ij}(Y)u_j \tag{A.30}$$

Hence,

$$\int\limits_0^\infty \iint\limits_\Omega l_i(t,a,X)\,da\,dX \to V/m \int\limits_\Omega [m_i(Y) \sum_{j=1}^n \gamma_{ij}(Y)u_j]\,dY \text{ at } t \to \infty. \tag{A.31}$$

The assertion (4.2) of Ergodic theorem follows from this relation.

*Remark.* One can show, using some known results on the rates of convergence in the renewal theorems, that for given model (4.1) under some conditions there exist such positive constants $c_1$, $c_2$ such that that for all $t\,|\,\Theta_i(t)/T_i - K\,|<c_1 \exp(-c_2\,t)$.

Next, let us suppose that some known conditions providing the possibility of limit transition under integral in the formula

$$Q_i(t) = \int\limits_\Omega \int\limits_0^\infty q(a,X)l_i(t,a,X)\,da\,dX$$

are fulfilled. For example, the transition is justified for a *finite* regulator when the function $q$ is integrated, bounded and equal to 0 outside the bounded domain in $[0. \infty)$ x $\Omega$, or for *bounded* regulators such that $Q_i(t) < const$ for all $t$, or under conditions of the theorem of Lebegue. Then

$$\lim_{t \to \infty} Q_i(t) = \lim_{t \to \infty} \int_{\Omega} \int_0^{\infty} q(a,X) l_i(t,a,X) da dX =$$

$$V/m \int_{\Omega} [q(a,X(a,Y)) m_i(Y) \sum_{j=1}^{n} \gamma_{ij}(Y) u_j] \mathrm{d}Y = V/m [TQ_i] \qquad \text{(see (4.4)),}$$

and the assertions (4.5), (4.6) of the Generalized Ergodic theorem for regulators $Q_i(t)$ are proved. The same assertions for regulators $G_i(a,t)$ can be proved similarly.

**Remark.** Using known results on the rates of convergence in the renewal theorems, one can show that for given model (4.1) under some conditions there exist such positive constants $c_1$, $c_2$ such that for all $t$

$$|Q_i(t)/[TQ_i] - K| < c_1 \exp(-c_2 t)$$
$$|G_i(a,t)/[TG_i(a)] - K| < c_1 \exp(-c_2 t).$$

At last, it is known that a distribution is uniquely defined by the values of all its (finite) regulators. Hence, from the validity of equation

$$\int_{\Omega} \int_0^{\infty} q(a,X) L(a,X) da dX = V/m \int_{\Omega} [q(a,X(a,Y)) m_i(Y) \sum_{j=1}^{n} \gamma_{ij}(Y) u_j] \mathrm{d}Y = V/m [TQ_i]$$

for all finite functions $q(a,X)$ it follows that the distribution $L(a,X)$ coincides with the distribution $l_i(a,X)$.

Hence, the reverse statement of the generalized ergodic theorem is valid: *if the relations (4.6) hold for all generalized variables then the succession system resides in the stationary (climax) state.*